\documentstyle[12pt]{article}

\begin{document}

\title{Noncritical M-theory: Three Dimensions}
\author{Michael McGuigan\\[.5cm]
Brookhaven National Laboratory\\[.5cm]
Upton, NY 11973\\[.5cm]
mcguigan@bnl.gov}
\date{}

\maketitle

\begin{abstract}
At the macrosopic level we study candidate 3D effective field theories associated to M theory in three dimensions. These represent analogs of 11D supergravity for eleven dimensional M-theory. At the microscopic level we study various  world volume (WV) theories that can be used to relate world sheet (WS) theories of two dimensional strings. These  are better defined for 3D targets than their 11D counterparts in part because there are no transverse dimensions. In particular M-theory on the target manifold $AdS^3$ and $R\times SL(2,R)/SO(2)$ are studied using a supermembrane and WV 3d gravity approach.
\end{abstract}

\newpage

\section*{I Introduction}

\indent
There are two main reasons to study a noncritical 3D version of M-theory. First it may tell us something about the 11D M-theory. Secondly, as we shall see, M-theory in 3D has the onshell degrees of freedom of 3D gravity and one matter field. This by itself is an unsolved problem. The situation is different from String theory in 2D which has the onshell degrees of freedom of  2D gravity and one matter field which is already renormalizable as a field theory, or string theory in 3D which contains an infinite tower of fields onshell.

Recently there has been a revival of interest in noncritical string theories in two dimensions [1,2]. This is largely due to the recent extension of the Matrix model to include fermions [1,2] as well as to include a wider variety of backgrounds such as D-branes [3,4] and blackholes [5,6]. Noncritical string theory in two dimensions is described macroscopically by two dimensional  dilaton gravity $(G_{IJ}^{(2)} ,\Phi )$
  plus a single matter field $T$
. Noncritical 2D string theory is distinguished from critical 10D string theory because it has no oscillator degrees of freedom on shell. Like critical 10D string theory noncritical string theory in two dimensions comes in several varieties, Type 0A [2], Type 0B [7], Type I , Type IIA [8], Type IIB [7], Heterotic $E8\times SO(8)$ [9] and Heterotic $SO(24)$ [9] . The macroscopic action has solutions of the form $R^2$ with linear dilaton, 2D blackhole, D0 brane and rolling $T$ solutions [10]. Noncritical two dimensional string theory at the microscopic level allows computations of vacuum energy, amplitudes for $T$ scattering and its effective action. Microscopic approaches to the theory include WS GS (Green-Schwarz) superstring action [11], WS NSR action [2], large N dual ($c=1$ matrix model approach) [1,2], conformal field theory (CFT2) approach with coset description SL(2,R)/SO(2) of a black hole background [12], and finally the superliouville and WV 2d supergravity approach [13].

Because of the existence of different noncritical two dimensional string theories one is interested in constructing 3D noncritical M-theory [11] to obtain a unified model of noncritical string theories. Before discussing the 3D theory let us review the familiar properties of 11D M-theory. At the macroscopic level 11D Critical M-theory is described by an effective field theory of eleven dimensional supergravity coupled to a three form potential [14]. Phenomenological considerations include compactification on $R^4 \times M^7$ with G2 Holonomy [15] or $R^4\times CY^6 \times S1/Z2$ with boundary gauge matter [16,17]. The main purpose of M-theory (whose exact formulation is unknown at this point) is to connect the five critical string theories Type IIA, IIB, Type I, Heterotic $E8\times E8$ and Heterotic $Spin(32)/Z2$ [14]. The effective theory has various solutions including M2, M5 brane solutions, blackhole solutions, $R\times T^{10}$, $R^3 \times CY^8$ wher $CY^8$ has $Spin(7)$ holonomy [18]. A viable microscopic critical M-theory would allow us to calculate vacuum energy, scattering amplitudes or  effective potentials and yield insight into the foundation of string models. At the microscopic level although we do not have a final formulation of M-theory, several approaches can be taken by analogy with string theory [19,20]. An early example was the world volume (WV) action of the supermembrane [21,22,23,24] that has manifest target space supersymmetry. More recently there is a large N dual approach for backgrounds of the form $S^5\times AdS^5\times S1/Z2$ [25] and a M(atrix) theory approach [26,27].
 
Based on the properties of noncritical string theory and the principle of double dimensional reduction (DDR)[28,29] where one target space and one world volume coordinate are compactified, one expects noncritical M-theory in three dimensions to be described macroscopically by a target space theory of three dimensional gravity plus a single matter field. Also based on these considerations noncritical 3D M-theory would be expected to contain no oscillator degrees of freedom as well as no dilaton. Again the main incentive to construct such 3D M-theory is to connect the two dimensional string theories Type 0A, type 0B, Type I, Type IIA, Type IIB, Heterotic $E8 \times SO(8)$ and Heterotic $SO(24$). The macroscopic theory is  expected to have solutions of the Black hole in 3 dimensions, $AdS^3$, $AdS^2 \times S^1$, and $R\times EAdS^2$. At the microscopic level one expects that a viable 3D M-theory would allow one to calculate vacuum energy, amplitudes and effective potentials as can be done in noncritical string theory. Microscopic approaches to noncritical M-theory in three dimensions are expected to be better defined than their eleven dimensional counterparts. This is mainly because there are no transverse target dimensions in 3D. In particular some microscopic approaches include a reparametrization invariant WV GS supermembrane with three target dimensions and a large N dual approach for targets $AdS^3$ with relation to Calogero-Moser model [11]. 

As one of the main motivations for studying 3D M-theory is to learn about the 11D theory, one may question whether what is learned in a better defined 3D context could be extrapolated to 11D. In a sense this is the same situation that exists already for 2D string models which are also in many ways very different from their 10D counterparts. In that case it has been found that 2D string models can still yield valuable insights. In particular we can point to the existence of topological phases of the 2D theory at high temperature [30] and the existence of exact solutions in the form a 2D black hole found in [12]. So that there is some expectation that one may also gain insight by studying a 3D microscopic version of M-theory.

This paper is organized as follows. In section II we describe macroscopic considerations of effective M-theoretic field theories. We consider effective field theories relating noncritical 3D M-theory to 2D 0A, IIA and Heterotic string effective actions. In section III we review some results from noncritical string theory to place noncritical M-theory in context. In section IV and V we discuss some M-theoretic reparametrization invariant WV actions related to the WS theories of noncritical  strings. In particular we consider the WV supermembrane action, the WV NSR membrane action and the WV 3d gravity approach, all with no oscillators and a single state. Finally in section VI we summarize the main results of the paper.

We will use the following notation:
$D$ denotes Target space dimension; $d$ world volume dimension; 
$\dot M,\dot N$
Target superspace indices; $M,N$
Target spacetime 11D or 3D indices; $\bar M,\bar N$
 transverse Target space indices; $I,J$
Target spacetime 10D or 2D indices; $\alpha ,\beta $
Target spinor indices; $\mu ,\nu $
3d World Volume indices; $i,j$
2d WS idices; $a,b$
$SL(2,R)$
group indices; $G$
Target space metric; $g$
World volume metric; 
$\Lambda  =  - \frac{1}{{\ell ^2 }}$ Target cosmological constant;
$\lambda  =  - \mu ^2 $ world volume cosmological constant; $A_{(n)} $
as a n form potential; and $F_{(n + 1)} $
as a n+1 form field strength. 

\section*{II Effective actions of  3D Noncritical M-theory}

\hspace{.5cm} Because noncritical M-theory is a largely unknown and unexplored topic it is helpful before delving into the 3D theory to first review some basic properties of 11D critical M-theory and the relation between critical 11D M-theory and 10D IIA supergravity [14]. 

For 11D  critical M-theory the spectrum of the effective 11D gravity is related to the degrees of freedom of 10D IIA effective action by [14]:

 \[
\begin{array}{l}
 (66,44,G_{MN} ) = (55,35,G_{IJ} ) \oplus (10,8,A_I ) \oplus (1,1,\Phi ) \\ 
 (165,84,A_{MNP} ) = (45,28,B_{IJ} ) \oplus (120,56,A_{IJK} ) \\ 
 (128,\Psi _M ) = (56 + 8,\Psi _I^{(1)} ) \oplus (56 + 8,\Psi _I^{(2)} ) \\ 
 \end{array}
\] 
\hfill (2.1)

\noindent In this notation $(N_f,N_d,F)$  represents $N_f$ number of field components and $N_d$ number of degrees of freedom for field $F$. The 11D fields are on the left and the 10D fields are on the right. Note how the 11D theory does not contain a dilaton. Instead one of the metric components is identified as the dilaton in the 10D theory. 

To go beyond the counting argument one can relate the effective action of the 11D supergravity coupled to a three form potential :
 
\[
I_M  = \int {d^{11} x} (\sqrt { - G} (R - (dA_{(3)} )^2  +  \ldots )
\]
\hfill (2.2)

\noindent to the IIA 10D effective theory of 10D supergravity coupled to a dilaton, two form potential, three form potential and vector field given by:

\[
I_{IIA}  = \int {d^{10} x} (\sqrt { - G} (e^{ - 2\Phi } (R + 4\left( {\nabla \Phi } \right)^2  + \left( {dB} \right)^2 ) + \left( {dA} \right)^2  + (dA_{(3)} )^2  \ldots )
\]
\hfill (2.3)

Following [14] we introduce the ansatz for the eleven dimensional metric where all fields are independent of $x_{11} $
:

\[ds_{11}^2  = G_{IJ}^{(10)} dx^I dx^J  + \eta ^2 (dx^{11}  + A_I dx^I )^2\] 
\[A_{(3)11IJ}  = B_{IJ} \]
\hfill (2.4)

\noindent Then on $M^{10}\times S^1$ the 11D action becomes

\[
\begin{array}{l}
 I = \int {d^{10} x} (\sqrt {-G^{(10)} } (\eta (R + (dA_{(3)} )^2 ) + \eta ^3 (dA)^2  + \eta ^{ - 1} (dB)^2  +  \ldots ) \\ 
  \\ 
 \end{array}
\]
\hfill (2.4)

\noindent Finally if one then rescales $G^{(10)}  = \eta ^{ - 1} G'$
 the action of the compactified theory is given by:

\[
I = \int {d^{10} x} (\sqrt { - G'} (\eta ^{ - 3} (R + 9\eta ^{ - 2} (d\eta )^2  + (dB)^2 ) + (dA)^2  + (dA_{(3)} )^2  +  \ldots )
\]
\hfill (2.5)

\noindent So that  the compactified 11D action agrees with IIA 10D effective action if the dilaton field is related to the to radius of extra dimension $\eta  = e^{2\Phi /3} $ [14].

\subsection*{Noncritical 3D M-theory relation to 2D 0A target space theory}

\indent
Now we shall try to repeat the same analysis for an effective action that could be used in principle to describe 3D M-theory. Note that the general problem of $3\rightarrow 2$ reduction starting from a 3D Target gravity and  obtaining 2D string inspired effective action is more complicated than simple scaling arguments and has been studied by Achucarro and Ortiz [31,32] and Cangemi [33].

Using a similar approach as above we search for an M-theory effective field theory that has the degrees of freedom of 2D 0A effective action (one massless scalar). To do so we introduce a macroscopic three dimensional effective field theory depending on the fields $G,A_{(2)} ,T$
that is characterized by no transverse oscillators, no dilaton, and a single scalar degree of freedom $T$. Also note there will no three form potential$A_{(3)}$  as it's field strengh $F_{(4)}$  would be a four form and vanish in 3D. To see that this Field content has the same degrees of freedom as the Type 0A theory, recall that the 2D 0A Target space theory is constructed from the GSO projection$(NS + ,NS + ),(NS - ,NS - ),(R + ,R - ),(R - ,R + )   $
[2]. Then the 0A theory  has only bosons in its spectrum with field content given by:
:

\[
\begin{array}{l}
 (NS,NS):\hspace{1.cm}    (G^{(2)} ,B,\Phi ),(T) \\ 
 (R,R):\hspace{1.3cm}         (A^{(1)} )(A^{(1')} ) \\ 
 \end{array}
\]
\hfill (2.6)

\noindent Metric fields, vector and antisymmetric fields have no oscillation degrees of freedom in two dimensions. Thus in 2D 0A has one massless field theoretic degree of freedom from the scalar $T$. Finally the 0A effective action is given by [2]:

\[
I_{0A}  = \int {d^2 x(\sqrt { - G^{(2)} } e^{ - 2\Phi } (R + \frac{8}{{\alpha '}} + 4(\nabla \Phi )^2  - \frac{1}{2}(\nabla T)^2  + \frac{1}{{\alpha '}}} T^2 )  + \frac{1}{4}(dA_{(1)} )^2  + \frac{1}{4}(dA_{(1')} )^2 )
\]
\hfill (2.7)

\noindent Note that like the 10D theory above the RR fields do not couple to the dilaton. Despite appearances $T$
 is massless after the rescaling $\tilde T = e^{ - \Phi } T$.

To see how the spectrum of the effective action for 3D M-theory compares to the spectrum of 2D 0A string theory look at the mapping of states given by:
\[
\begin{array}{l}
 (6,0,G_{MN} ) = (3, - 1,G_{IJ} ) \oplus (2,0,A_{(1)I} ) \oplus (1,1,\Phi ) \\ 
 (3,0,A_{(2)MN} ) = (1,0,B_{IJ} ) \oplus (2,0,A_{(1')I} ) \\ 
 (1,1,T) = (1,1,T) \\ 
 \end{array}
\]
\hfill (2.8)

\noindent Again we use the notation $(N_f ,N_d ,Field)$
 to illustrate the mapping. Note for example that the 3D metric has zero degrees of freedom and the 2D metric and dilaton on the right have the necessary degrees of freedom match that. The 2D metric has -1 degrees of freedom because it has two reparametriztion symmetries and one space-space component. Together with the dilaton it has the degrees of freedom to cancel two ghosts in two dimensions associated with 2D reparametrization invariance. 

The correspondence can be considered more closely by considering the 3D effective action:
\[
I_M  = \int {d^3 x(\frac{1}{{16\pi G_N }}E(R - 2\Lambda )}  + \frac{1}{6}EF_{(3)\mu \nu \lambda } F_{(3)} ^{\mu \nu \lambda }  + \frac{1}{2}EG^{\mu \nu } \partial _\mu  \tilde T\partial _\nu  \tilde T)
\]
\hfill (2.9)

\noindent Throughout this paper we will set $16\pi G_N  = 1$.
In 3D one has the special situation where a scalar field has the same number of degrees of freedom as a vector field. Thus there is an alternative representation given by a 3D metric $G$ and vector field $A$ with a mapping onto 2D 0A fields: 

\[\begin{array}{l}
 (6,0,G_{MN} ) = (3, - 1,G_{IJ} ) \oplus (2,0,A_{(1)I} ) \oplus (1,1,\Phi ) \\ 
 (3,1,A_M ) = (1,1,T) \oplus (2,0,A_{(1')I} ) \\ 
 \end{array}\]\hfill (2.10)

\noindent In this case the alternative effective action for 3D M-inspired field theory is:
\[
I_M  = \int {d^3 x(E(R - 2\Lambda )}  + \frac{1}{4}EF_{MN} F^{MN} )
\]
\hfill (2.11)

\noindent Where $G_{MN}  = E_M^a E_N^a $
 and $E = \det (E_M^a )$
. For the most part we will use the effective action involving the first representation with the T field however.

The equations of motion following from the first representation (2.9) involving $G$, $F_{(3)}  = dA_{(2)}$ and $T$ have the form:

\[R_{MN}  - \frac{1}{2}RG_{MN}  = T_{MN}  - \Lambda G_{MN}\]\vspace{-.2cm}
\[ \nabla _M F_{(3)NP}^M  = 0\]\vspace{-.1cm}
\[ \nabla ^2 T =V^\prime(T)\]
\hfill (2.12)

\noindent
Where the stress tensor of the matter fields is given by:

\[T_{MN}  = \frac{1}{{12}}G_{MN} F_{(3)} ^2  - F_{(3)M}^{  PQ} F_{(3)NPQ}  + \nabla _M T\nabla _N T - \frac{1}{2}G_{MN} ((\nabla T)^2  - \frac{1}{2}V(T))\]
\hfill (2.13)

\noindent
and $V(T)$ is an effective potential for the $T$ field.

To relate to 3D effective theory to 2D 0A string theory we parametrize the solution to (2.12) as:

\[ds_3^2  = ds_2^2  + \eta ^2 (du + A_ +  dt^ +   + A_ -  dt^ -  )^2 \]
\[A_{(2)I3}  = A_I \]
\[T = \tilde T\]
\hfill (2.14)

\noindent
Equations (2.12) have been studied in [34]. They have solutions of the form:

\[ ds_3^2  = \ell ^2 \frac{{(dt^ +   - dt^ -  )^2 }}{{4(t^ +   - t^ -  )^2 }} + \ell R_1 \frac{{du(dt^ +   + dt^ -  )}}{{(t^ +   - t^ -  )}} + R_1^2 du^2\] 
\[ F_{(3)MNP} dx^M dx^N dx^P  = \frac{1}{{2(t^ +   - t^ -  )^2 }}dt^ +  dt^ -  du\]  
\[ T = 0\] 
\hfill (2.15)

\noindent where the two dimensional metric, vector field  and third dimension radius are given by:

\[ds_2^2  =  -\ell ^2 \frac{{dt^ +  dt^ -  }}{{(t^ +   - t^ -  )^2 }} \]
\[ A_ \pm   = \frac{1}{{2(t^ +   - t^ -  )}}\]
\[ \eta ^2  = R_1^2 \ell ^{ - 2} \] 
\hfill (2.16)

\noindent The solutions (2.16) compare with those of 0A string theory on $AdS^2$ in the near horizon limit where the gradient of the dilaton field is much smaller than other terms in the action. Variation of the 0A action yields the 0A equations of motion: 
\[R - 4(\nabla \Phi )^2  - 4\nabla ^2 \Phi  + \frac{8}{{\alpha '}} = 0\]  
\[\partial _ \pm  ( \frac{{F_{ +  - } }}{{\sqrt { - G} }}) = 0\] 
\[4e^{ - 2\Phi } (\nabla \Phi )^2  + \frac{{16}}{{\alpha '}}e^{ - 2\Phi }  - F^2  = 0\]
\hfill (2.17)

The asumption that $A$ and $A^\prime$ fluxes are equal is made to cancel dilaton tadpoles [34,35,36]. The near horizon solution to the 0A effective action is given by the condition $(\nabla \Phi )^2  = 0$. In that case solutions to () iare given by [34,35,36]:.

\[ds_2^2  =  - \ell ^2 \frac{{dt^ +  dt^ -  }}{{(t^ +   - t^ -  )^2 }}\] 
\[ A_ \pm   = \frac{q}{{2(t^ +   - t^ -  )}}\]
\[ e^{ - 2\Phi }  = e^{ - 2\Phi _0 }\] 
\hfill (2.18)

\noindent So there is agreement with the solution (2.16) from the 3D effective action with that from the 0A action in the near horizon limit.
\pagebreak

\subsection*{3D Target Supergravity and 2D Target IIA Supergravity}

\indent For the 3D effective action related to the IIA 2D supersymmetric string theory the analogy is closer to the 11D case as the target space spectrum is supersymmetric. Following the same procedure as above the spectrum matching from 3D to 2D is given by:

\[\begin{array}{l}
 (6,0,G_{MN} ) = (3, - 1,G_{IJ} ) \oplus (2,0,A_I^{(1)} ) \oplus (1,1,\Phi ) \\ 
 (6,0,\Psi _M ) = (4, - 1,\Psi _I ) \oplus (2,1,\lambda ) \\ 
 (1,1,\Upsilon ) = (1,1,\Upsilon ) \\ 
 (1,1,T) = (1,1,T) \\ 
 \end{array}\]
\hfill (2.19)

\noindent where $G$ is the 3D metric, $\Psi$ is the 3D Rarita Schwinger field, and $\Upsilon$ is the superpartner to the $T$ field.

The 3D macroscopic effective action in this case is written:

\[
\begin{array}{l}
 I_M  = \int {d^3 x}\Big( \varepsilon ^{MNP} (E_{Ma} R_{NP}^a  + \frac{1}{2}\bar \Psi _M (D_N \Psi _P  - D_P \Psi _N ) - \frac{2\Lambda }{3}\varepsilon _{abc} E_M^a E_N^b E_P^c  \\[.2cm] 
\hspace{.5cm}  + i\sqrt \Lambda  \frac{1}{2}E_M^a \bar \Psi _N \Gamma _a \Psi _\P  ) + E((E^a E_a^T )^{ - 1MN} \partial _M  T\partial _N T + (E^{ - 1} )^{Ma} \bar \Upsilon \Gamma _a \partial _M  \Upsilon \Big) \\ 
 \end{array}
\]
\hfill (2.20)

\noindent The bosonic part of the IIA 2D Target theory is in turn described by the effective action [8]:

\[
\begin{array}{l}
 I_{IIA}  = \int {d^2 } x\sqrt { - G^{(2)}} (e^{ - 2\Phi } (R + 4(\nabla \Phi )^2  + \frac{8}{{\alpha '}} - (\nabla T)^2 ) - \frac{1}{4}F^2 ) \\ 
                   \\ 
 \end{array}
\]
\hfill (2.21)

\noindent Again one can use the ansatz:

\[
ds_{(3)}^2  = G_{IJ}^{(2)} dx^I dx^J  + \eta ^2 (dx^3  + A_I dx^I )^2 
\]
\hfill (2.22)

\noindent and proceed as above to compare solutions to the classical solutions from (2.20) and (2.21). The main new feature is the space time supersymmetry of (2.20) as well as the fact the 2D theory only contains a single vector field.
\pagebreak

\subsection*{3D M-theory effective action and the 2D Heterotic $E_8 \times SO(8)$ String}

\indent
The 3D M-inspired effective action reduction to the 2D Heterotic $E_8\times SO(8)$ is in some ways the most physically interesting case because it represents  3D to 2D analog of the 11D to 10D reduction on $S^1/Z_2$ for Heterotic M-theory. The macroscopic theory can be represented as a 3D effective action with 2D boundary matter transforming as $E_8$ or $SO(8)$ respectively on 2D target space boundaries.$\partial _1 M$
or $\partial _2 M$
. The effective action can be written in the downstairs (interval) representation as

$$I_M  = \int {d^3 x} E(R - 2\Lambda ) + \int\limits_{\partial _1 M} {d^2 x(F^2  + (\nabla T)^2) }  + \int\limits_{\partial _2 M} {d^2 x(F^{'2}  + (\nabla T^\prime)^2 )}$$
\hfill (2.23)

\noindent Here $T$ and $T^{\prime}$ are scalars tranforming in the fundamental representation of $E_8$ and $SO(8)$ respectively. For comparison the 2D Heterotic string effective action is given by

\[I_{Het}  = \int {d^2 x\sqrt { - G^{(2)}} } e^{ - 2\Phi } (R + 4(\nabla \Phi )^2  + \frac{8}{{\alpha '}} - \frac{1}{4}F^2  - \frac{1}{4}F'^2  + (\nabla T^{(2)})^2 )\]
\hfill (2.24)

\noindent Similar to the previous cases the dimensional reduction on $S^1/Z_2$ or interval is obtained using the ansatz:

\[
ds_3^2  = \eta ^2 G^{(2)}_{IJ} dx^I dx^J  + \eta ^2 dx_3^2 
\]
\hfill (2.25)

\noindent Here the vector usually obtained from the 3D gravity metric is projected out by $S^1/Z_2$ symmetry as $G_{3i} $
is odd under the symmetry $x^3  \to  - x^3 $
[17]. Now if we set:

	\[
\begin{array}{l}
 \eta ^2  = e^{ - 2\Phi }  \\ 
 T=T' = e^{ - \Phi } T^{(2)} \\ 
 F = F \\ 
 F' = F' \\ 
 \end{array}
\]
\hfill (2.26)

\noindent we obtain agreement between the dimensionally reduced 3D M-inspired effective action and the 2D effective action of (2.24) after setting $\Lambda  =  - \frac{8}{{\alpha '}}$
. The 2D Heterotic effective action has charged black hole solutions [9] and 2D coset constructions [37] however no matrix model has yet been found. 
\pagebreak

\subsection*{Solutions to M-inspired 3D effective actions}

\indent
The M-inspired 3D effective actions (2.9,2.20,2.23) lead to  interesting solutions that have been studied previously in the context of 3D gravity. Three cases are of particular interest. The first is the 3D black hole. 

\subsection*{BTZ Black Hole}

\indent
Setting the cosmological constant to  $\Lambda  =  - \frac{1}{{\ell ^2 }}$
 the 2+1 Black hole solution of Banados, Teitelboim and Zannelli (BTZ) [38] can be written:

\[ds_3^2  =  - (\frac{{r^2 }}{{\ell ^2 }} - M)dt^2  + (\frac{{r^2 }}{{\ell ^2 }} - M)^{ - 1} dr^2  + r^2 d\phi ^2 \]
\hfill (2.27)

\noindent The BTZ blackhole can be constructed as a region of $AdS^3$ with identification of its boundaries to take into account the perioidicity of the angular coordinate $\phi $
. It has a generalization to rotating and charged 2+1 black holes. For the simple Schwarzschild like case (2.27) the mapping to a region of $AdS^3$ is for $r > \ell ^2 M$

\[\begin{array}{l}
 x = \cosh (Mt)\exp (\ell M\phi ) \\ 
 y = \sinh (Mt)\exp (\ell M\phi ) \\ 
 z = \exp (\ell M\phi ) \\ 
 \end{array}\]
\hfill (2.28)

\noindent so the metric becomes for $z > 0$

\[ds_3^2  = \frac{{\ell ^2 }}{{z^2 }}(dx^2  - dy^2  + dz^2 )\]
\hfill (2.29)

\noindent with identifications:$(x,y,z) \sim (e^{2\pi \ell M} x,e^{2\pi \ell M} y,e^{2\pi \ell M} z)$. Now defining coordinates $\gamma  = x + y,\bar \gamma  = x - y,z = e^{ - \varphi } $
the metric is written $ds_3^2  = \ell ^2 (e^{2\varphi } d\gamma d\bar \gamma  + d\varphi ^2 )$
 with indentifications  $(\gamma ,\bar \gamma ,\varphi ) \sim (e^{2\pi \ell M} \gamma ,e^{2\pi \ell M} \bar \gamma ,\varphi  - 2\pi \ell M)$
. We will return to this parametrization in section IV when we discuss a microscopic formulation of the theory. The metric also admits Killing spinors and represents a solution to the supersymmetric 3D action (2.20)[39,40].

\subsection*{Strominger $AdS^3  \to AdS^2 $ Solution}

The second solution to (2.20) is the reduction  $AdS^3$ to $AdS^2$ solution of Strominger [34] given by 
	
	\[
\begin{array}{l}
 ds_3^2  = ds_2^2  + \ell ^2 \eta ^2 (dx^3  + A_t dt + A_2 dx^2 )^2  \\ 
 \eta ^2  = \frac{{R_1^2 }}{{\ell ^2 }} + \frac{{R_2^2 U^2 }}{{\ell ^4 }} \\ 
 A_t  = \frac{{R_1^2 }}{{R_2 \ell ^2 }}\eta ^{ - 2}  \\ 
 ds_2^2  = \frac{{ - R_2^2 U^4 }}{{\ell ^4 R_1^2  + \ell ^2 R_2^2 U^2 }}dt^2  + \frac{{\ell ^2 }}{{U^2 }}dU^2  \\ 
 \end{array}
\]
\hfill (2.30)

\noindent where the reduction is $(t,U,x_3 ) \to (t,U)$. The spacetime reduces to $AdS^2$ in the near horizon limit $U^2 R_2^2 \ell ^{ - 4}  \to 0$
while the radius of the third dimension $\eta $
takes a constant value. This space can also be represented by the Poincare form (>) but with identifications $(\gamma ,\bar \gamma ,\varphi ) \sim (e^{4\pi T} \gamma ,\bar \gamma  + 2\pi R_2 ,\varphi  - 2\pi R_1 )$
. The near horizon solution with $R_2  = 0$
 was used above to compare the solutions of 3D M-theory effective actions and the Type 0A solutions effective action. 

\subsection*{Coussaert-Henneaux (CH) solution with topology $AdS^2  \times S^1$}

\indent 
The last solution we want to consider is Coussaert-Henneaux (CH) [41] solution which is given by the metric:

\[ds^2  = \frac{{\ell ^2 }}{4}( - dt^2  + dr^2  + 2\sinh (r)dtd\phi  + d\phi ^2 )\]
\hfill (2.31)

\noindent with the identification $\phi  \sim \phi  + 2\pi $
. This solution has topology $AdS^2 \times S^1$ that differs from that of the 3D black hole. In term of the parametrization:

\[ds_3^2  = ds_2^2  + \eta ^2 (d\phi  + A_t dt)^2 \]
\hfill (2.32)

\noindent The solution is of the form:
\[\begin{array}{l}
 ds_2^2  =  - \frac{{\ell ^2 }}{4}\cosh ^2 (r)dt^2  + dr^2  \\ 
 A_t  = \sinh (r) \\ 
 \eta  = \frac{\ell }{2} \\ 
 \end{array}\]
\hfill (2.33)

\noindent The CH solution has the property that $(\nabla \eta )^2  = 0$
while the 2D curvature tensor obeys $R^{(2)}  =  - \frac{8}{{\ell ^2 }}$
and the two dimensional metric describes $AdS^2$. In the presence of the $F_{(3)}$ field the solution has a non zero $A_{(2)}$ field given by $A_{(2)\phi t}  = \frac{\ell }{2}\sinh (r),A_{(2)\phi r}  = A_{(2)tr}  = 0$
in accordance with the identification (2.8) and with $F_{(3)MNP}  = \frac{2}{\ell }\varepsilon _{MNP} ( - G)^{1/2} $
. In addition the solution possesses two Killing spinors and preserves supersymmetry in the target space. More information and classification of these and other solutions to (2.12) can be found in [42,43] and [44].

All three of these solutions solve the equations of motion of the 3D M-inspired effective actions at tree level. However one needs to investigate the microscopic realization of noncritical M-theory in order to see whether they are still solutions at the quantum level. This is the subject to which we now turn.
\pagebreak

\section*{III Brief Review of Noncritical  String Theory}

\indent
To set the stage for noncritical M-theory we briefly review some aspects of  noncritical string models for later comparison. 

\subsection*{GS Superstring on $AdS^2$} 

First consider the WS IIA GS String on $AdS^2$ recently considered by Verlinde in [11]. Its action can be written as a sigma model 

\[
 S_{GS}  = \int {d^2 \sigma ( - \frac{1}{2}} \sqrt { - g} T(Z) + \frac{1}{2}\sqrt { - g} g^{ij} \partial _i Z^{\dot I} \partial _j Z^{\dot J} E_{\dot I}^A E_{\dot J}^A  + \frac{1}{2}F_{\dot I\dot J} \varepsilon ^{ij} \partial _i Z^{\dot I} \partial _j Z^{\dot J}) 
\]
\hfill (3.1)

\noindent with $T,E_I^A ,F_{IJ} $
 background fields. For the specific near Horizon background of $AdS^2 $
 and $T = 0$ the action is given a $Osp(2|1;R)/SO(2)$ supercoset sigma
model. The WS action  is written as:

\[
S_{GS}  =  - \int {d^2 \sigma } \frac{{\left( {\bar \partial Z + \theta \bar \partial \theta } \right)\left( {\partial \bar Z + \bar \theta \partial \bar \theta } \right)}}{{\left( {Z - \bar Z - \theta \bar \theta } \right)^2 }}\]
\hfill (3.2)

So that the action reduces to a  sigma model with metric and Flux target space backgrounds:

\[
 ds_2^2  =  - \frac{1}{4}\frac{{dZd\bar Z}}{{(Z - \bar Z)^2 }} = \frac{{ - dX_1^2  + dX_2^2 }}{{X_2^2 }}
\]
\[ 
 E_{\dot I}^1 dZ^{\dot I}  = \frac{{dZ + \theta d\theta }}{{Z - \bar Z - \theta \bar \theta }} 
\]
\[
 E_I^2 dZ^I  = \frac{{d\bar Z + \bar \theta d\bar \theta }}{{Z - \bar Z - \theta \bar \theta }}
\]
\[
 F_{IJ} dZ^I dZ^J  = \varepsilon _{AB} E_I^A E_J^B dZ^I dZ^J   
\]
\hfill (3.3)

\noindent Verlinde [11] has studied the spectrum generating algebra for this model using the coset supergroup sigma model currents and $\kappa $
symmetry to show that the theory contains no transverse oscillations on $AdS^2 $. Further discussion of noncritical GS superstrings can be found in [45].
\pagebreak

\subsection*{NSR spinning string in two Target dimensions}

\indent
Next we consider the WS NSR 0A 2D string with a Liouville action [2]. In the 2D case the Liouville mode is taken as a space dimension. The WS NSR action can then be written using a Liouville superfield as:

\begin{eqnarray*}
 S_{Liouville}  &=& \int {d^2 \sigma d^2 \theta (\frac{1}{2}} D_\theta  \Phi _L D_{\bar \theta } \Phi _L  + 2\mu \exp (\Phi _L /2) + \frac{1}{2}D_\theta  XD_{\bar \theta } X) \\ 
 & =& \int {d^2 \sigma (\frac{1}{2}(\partial \phi )^2  - \frac{i}{2}\bar \zeta \partial \zeta  + \frac{1}{2}\mu ^2 e^\phi   - \mu e^{\phi /2} \bar \zeta \zeta  + \frac{1}{2}(\partial X)^2  - i\psi \partial \psi )} \\ 
 & =& S_{L - Kinetic}  + \int {d^2 \sigma (\frac{1}{2}\mu ^2 e^\phi   - \mu e^{\phi /2} \bar \zeta \zeta  + \frac{1}{2}(\partial  X)^2  - i\psi \partial \psi )} \\ 
 \end{eqnarray*}
\hfill (3.4)

\noindent where we have expanded out the superspace components. 

This form of the action  can be related to a to WS sigma model which is expressed as :

\[\begin{array}{l}
 S_{sigma}  = \int {d^2 \sigma (\sqrt { - \hat g} } \hat g^{ij} \partial _i X^I \partial _j X^J G_{IJ} + \sqrt { - \hat g} T(X^I ) + \sqrt { - \hat g} \mathord{\buildrel{\lower3pt\hbox{$\scriptscriptstyle\frown$}} 
\over R} \Phi (X^I )) - \frac{i}{2}\bar \zeta \partial \zeta  - \mu e^{\phi /2} \bar \zeta  \zeta  - i\psi \partial \psi)  \\ 
 \end{array}\]
\hfill (3.5)

\noindent where $X^I  = (\phi ,X)$
with the identification of 2D target space backgrounds: 

\[\begin{array}{l}
 ds_2^2  = d\phi ^2  + dX^2  \\ 
 T = \mu ^2 e^\phi   \\ 
 \Phi  = \phi  \\ 
 \end{array}\]
\hfill (3.6)

\noindent In (3.5) we have allowed for a term related to the curvature of the World Sheet. These background fields solve the equations of motion from the 2D Target space theory (2.7).
\pagebreak

\subsection*{WS 2d gravity approach to Noncritical String Theory}

\indent 
A third action related to the NSR and sigma model form is the WS 2d supergavity action [46,47] coupled to single scalar multiplet $(X,\psi )$
. It is related the NSR Liouville action in the conformal gauge. The action is written as:

\[\begin{array}{l}
 S_{2dgrav}  = S_{L - Kinetic}  - \frac{1}{2}\int {d^2 \sigma (\mu ^2 \varepsilon _{ab} \varepsilon ^{ij} e_i^a e_j^b  - 2\mu \varepsilon ^{ij} \bar \chi _i \chi _j  + \sqrt { - g} (g^{ij} \partial _i X\partial _j X - i\bar \psi (e^{ - 1} )^{ai} \gamma _a \partial _i \psi } ) \\[.4cm] 
\hspace{1.0cm}  - \frac{1}{2}\int {d^2 \sigma \sqrt { - g} (\bar \chi _i (e^{ - 1} )^{jb} (e^{ - 1} )^{ia} \gamma _a \gamma _b \psi \partial _j X + \frac{1}{4}\bar \psi \psi \bar \chi _i (e^{ - 1} )^{jb} (e^{ - 1} )^{ia} \gamma _a \gamma _b \bar \chi _j })  \\ 
 \end{array}\]
\hfill (3.7)

In the above action $\chi_i$ is the two dimensional Rarita-Schwinger field. The conformal gauge is given by:

\[
\begin{array}{l}
 g_{\alpha \beta }  = e^\phi  \mathord{\buildrel{\lower3pt\hbox{$\scriptscriptstyle\frown$}} 
\over g} _{\alpha \beta }  = (e^{\phi /2} \mathord{\buildrel{\lower3pt\hbox{$\scriptscriptstyle\frown$}} 
\over e} _\mu ^a )(e^{\phi /2} \mathord{\buildrel{\lower3pt\hbox{$\scriptscriptstyle\frown$}} 
\over e} _\nu ^a ) \\ 
 \chi _\mu   = e^{\phi /2} \mathord{\buildrel{\lower3pt\hbox{$\scriptscriptstyle\frown$}} 
\over e} _\mu ^a \gamma _a \zeta  \\ 
 \end{array}
\]
\hfill (3.8)

\noindent and the Liouville kinetic term in the WS 2d gravity approach arises as a one-loop correction to the effective action. 

All three actions (3.4,3.5,3.7) agree if the number of target space dimensions is given by:

\[D = N_L  + N_S  = \left( {3 - 2} \right) + N_S  = 1 + N_S  = 2\]
\hfill (3.9)

\noindent where $N_L$ is the number of Liouville fields (3 metric components minus two reparametrization symmetries) and $N_S$ is the number of scalar superfields present. 

The fact that the NSR , sigma model and WS 2d gravity representations are equivalent means that the path integral based on these actions gives the same physical results for vacuum energy, scattering amplitudes, etc. This means that:

\[\begin{array}{l}
 Z_{2dgrav} (WS) = \int {DeDbDcD\chi D\beta D\gamma DXD\psi e^{iS_{grav} } }  \\[.4cm]  
  = Z_{sigma} (WS) = \int {DX_1 DbDcD\psi _1 D\beta D\gamma DXD\psi e^{iS_{sigma} } }  \\[.4cm] 
  = Z_{Liouville + matter} (WS) = \int {D\phi DbDcD\zeta D\beta D\gamma DXD\psi e^{iS_{Liouville} } }  \\ 
 \end{array}\]
\hfill (3.10)

\noindent are identical once a GSO projection has been chosen. In these expressions $b,c$ represent reparametrization ghosts. In particular for the 0A 2D noncritical theory the GSO projection is $(NS+,NS+)(NS-,NS-)(R,R)$. For this GSO projection the one loop string correction to the target space cosmological constant can explicitly be evaluated as the path integral over $T^2$. In [2] it was shown to be:

\[Z_{0A} (T^2 ) =  - \frac{{\ln (|\mu |)}}{{12\sqrt 2 }}(2\frac{R}{{\sqrt {\alpha '} }} + \frac{{\sqrt {\alpha '} }}{R}) = -\int\limits_F {d^2 \tau \frac{1}{{\tau _2^2 }}} Z_{SL} Z_X Z_{ghost} \]
\hfill (3.11)

\noindent where:

\begin{eqnarray*}
 Z_X &=& \frac{R}{{\sqrt {\alpha '} }}\tau _2^{ - 1/2} |D_{r,s} |^2 \sum\limits_{mn} {e^{ - S_{m,n} } }  \\ 
 Z_{SL} &=& \frac{{V_L }}{{2\pi \sqrt 2 }}\tau _2^{ - 1/2} |D_{r,s} |^2  \\ 
 Z_{ghost}  &=& \frac{1}{2}\tau _2^{ - 1} |D_{r,s} |^{ - 4}  \\ 
 \end{eqnarray*}
\hfill (3.12)

\noindent and $D_{rs}$ is the determinant of the scalar laplacian on $T^2$ with $r,s$ boundary conditions (periodic or antiperiodic in $\sigma$ or $\tau$), $S_{m,n}$ is the action for winding and momentum modes. The path integral reduces to an integral over moduli of $T^2$.

The main lessons that we wish to take away from this brief review of noncritical microscopic 2D string theory are: in all cases reparametrization Ghosts cancel any Liouville or scalar oscillators (see the cancellation of determinants in (3.11) so there are no transverse oscillations. The spectrum has no oscillators and one physical state $T$. Finally the path integral definition of Target space cosmological constant reduces to integral over moduli.
 
\section*{IV Supermembrane with $AdS^3$ Target Space}

\indent
Historically one of the first approaches tried for a microscopic description of M-theory was to replace the World Sheet of string theory by some sort of World Volume theory. In this paper we use the term World Volume theory in a more general sense than typically used in, for example, a membrane or supermembrane approach. We define a world volume theory as any microscopic reparametrization invariant action that reduces to a world sheet string microscopic action under the process of double dimensional reduction. This reduction could be to a GS, NSR or even 2d gravity formulation of the world sheet string theory. A supermembrane action in our sense is a particular case of this, a specific world volume action that has the additional property that it directly reduces to the WS GS formulation of string theory under double dimensional reduction. 

The WV approach to M-theory has some attractive features. After all the WS representation of string theory united world lines that look very different. For example in the world sheet approach the graviton or photon world lines are treated together. This unification is at the heart of the successful quantization of gravity by string theory in which separate particles are represented by different nodes of a single string. If a WS theory could relate world lines that look different, then a WV theory could certainly unite WS theories that look different, for example Heterotic and Type I string theories could be represented by different limits of the same world volume. In this way a WV formulation of M-theory could unite all the seemingly different 10D string models. Also, already in string theory, even before the advent of M-theory, finite WV Chern Simons theories (CFT3s) were used to unite the many two dimensional conformal field theories (CFT2s) that describe large numbers of string vacua in the process termed "taming the conformal zoo" [48]. So there was a precedent that a thickening of the world sheet would be involved in a deeper understanding of string models.

The world volume approach to 11D M-theory suffered from many problems however. Most importantly the WV theories were nonrenormalizable, as typically then are written as a type 3d sigma model or a three dimensional version of the nambu action. Also these models suffered from a continuous spectrum [49], they have a complicated vertex operator in comparison with string models [50], finally they were found to be unstable, both with respect shape perturbations (spikes), and topology change (no dilaton) [51]. Although the shape perturbations have been given a multiparticle interpretation [52] the other problems are still present in the 11D theory.

For a microscopic 3D M-theory that we study in this paper the situation is somewhat better however. As the target space for 3D M-theory is the same dimension of the microscopic WV theory in a sense there are no transverse dimensions generating an infinite tower of onshell states. This means there will be no transverse oscillator modes present and effectively one expects only one physical state in the theory. This is the same situation as found for string theory with two target space dimensions. Actually the property that both 3D M-theory and 2D string theory have only finite number of states is important if 3D M-theory is to produce some type of 2D string model under the process of double dimensional reduction. So for 3D M-theory one can proceed further in the world volume approach which will be better defined than in the 11D theory. 

In this and the next section we will consider different continuous approaches to microscopic 3D M-theory. We will consider a WV GS type Membrane theory with target space $AdS^3$. Next we will consider WV NSR type spinning Membrane. Finally we will study a WV 3d gravity approach to noncritical M-theory with target space $R \times SL(2,R)/SO(2)$. All these theories will have the property of no oscillators and a discrete set of states.
\pagebreak

\subsection*{GS WV Action}

First we consider a WV GS type Membrane with target $AdS^3$. This is a direct generalization of Verlinde's treatment of the WS GS string on $AdS^2$ [11] and is written by the WV action:

\begin{eqnarray*}
 S_{SM}  = \int {d^3 \sigma \Big( - \frac{1}{2} \sqrt { - g} T(Z)} 
+ \frac{1}{2}\sqrt { - g} (g^{\mu \nu } \partial _\mu  Z^{\dot M} \partial _\nu  Z^{\dot N} E_{\dot M}^A E_{\dot N}^A \\
+ \frac{1}{2}\varepsilon ^{\mu \nu \rho } \partial _\mu  Z^{\dot M} \partial _\nu  Z^{\dot N} \partial _\rho  Z^{\dot P} E_{\dot M}^A E_{\dot N}^B E_{\dot P}^C \varepsilon _{ABC}\Big)
\end{eqnarray*} 
\hfill (4.1) 

\noindent The world volume scalar $Z^M $
parametrizes a target superspace with coordinates $Z^{\dot M}  = (X^M ,\theta ^\alpha  )$.
The action has invariance under world volume reparametrizations. 

In general for an arbitrary extended object with world volume dimension d and Target space dimension D a gauge choice exists which splits the transverse field theoretic degrees of freedom from longitudal zero modes. It is given by $X^\mu  (\sigma ) = \sigma ^\mu  $
 for $X^M (\sigma ) = (X^\mu  (\sigma ),X^{\bar M} (\sigma ))$
 with $\mu  = 1,2, \ldots ,d$
and $\bar M = d + 1, \ldots ,D$
. The number of onshell scalar bosonic degrees of freedom is then given by $N_S  = D - d$
. For the case of the supermembrane we have $d = 3$
, so that:
\[D = 3 + N_S \]
\hfill (4.2)

\noindent Now for $D = 3$
considered in this paper we have $N_S  = 0$
so there are no transverse degrees of freedom and classically the Supermembrane is completely specified by the zero modes $X^\mu  (\sigma ).$

The general sigma model discussion of reference [22]: 

\[ S_{SM}  = \int {d^3 \sigma \Big( - \frac{1}{2} \sqrt { - g}  + \frac{1}{2}\sqrt { - g} g^{\mu \nu } \partial _\mu  Z^{\dot M} \partial _\nu  Z^{\dot N} E_{\dot M}^A E_{\dot N}^A + \frac{1}{2}\varepsilon ^{\mu \nu \rho } \partial _\mu  Z^{\dot M} \partial _\nu  Z^{\dot N} \partial _\rho  Z^{\dot P} A_{\dot M\dot N\dot P}\Big)} \]
\hfill (4.3)

\noindent applies to the 3D supermembrane action (4.1) with the background $T = 1$
and $A_{MNP}  = \varepsilon _{abc} E_M^a E_N^b E_P^c $
. The general supermembrane action admits a flat background solution only for dimension 4, 5, 7 and 11. However here we are interested in a curved background so that $D=3$ is admissible.
\pagebreak

Kappa symmetry of the sigma model in a supersymmetric background is given by [22]:

\[
\begin{array}{l}
 \delta Z^{\dot M} E_{\dot M}^a  = 0 \\ 
 \delta Z^{\dot M} E_{\dot M}^\alpha   = (1 + \Gamma )_\beta ^\alpha  \kappa ^\beta   \\ 
 \end{array}\]
\hfill (4.4)

\noindent where $\Gamma $
Gamma is defined by:

\[\Gamma  = \frac{1}{6}\frac{1}{{\sqrt { - g} }}\varepsilon ^{ijk} \varepsilon ^{MNP} E_M^a E_N^b E_P^c \Gamma _{abc} \partial _i Z^M \partial _j Z^N \partial _k Z^P \]
\hfill (4.4)

\noindent and $\Gamma _{abc} $
is the antisymmetrized product of three gamma matrices. The Kappa invariance eliminates the two fermionic field theoretic degrees of freedom. The local world volume invariance eliminates the three bosonic field theoretic degrees of freedom so that only the ground state is physical. 

\subsection*{ $OSp(2|1;R)$  3d sigma model}

Following Verlinde [11] we can write the WV action as a sigma model over the supergroup manifold $OSp(2|1;R)$.  This supergroup is described by  three bosonic and two fermionic fields and can be parametrized as [11,40,53]: 

\[g_{OSp}  = \left( {\begin{array}{*{20}c}
   {zf^{ - 1/2} w^{1/2} } & {\bar zf^{ - 1/2} w^{ - 1/2} } & { - z\bar \theta f^{ - 1}  + \bar z\theta f^{ - 1} }  \\
   {f^{ - 1/2} w^{1/2} } & {f^{ - 1/2} w^{ - 1/2} } & {\theta f^{ - 1}  - \bar \theta f^{ - 1} }  \\
   {\theta f^{ - 1/2} w^{1/2} } & {\bar \theta f^{ - 1/2} w^{ - 1/2} } & {1 + \bar \theta \theta f^{ - 1} }  \\
\end{array}} \right)\]
\hfill (4.5)

\noindent with $f = z - \bar z + \bar \theta \theta$. The bosonic part of the supergroup is $SL(2,R)\sim AdS^3$ contained in the upper left $2\times 2$ matrix. The inverse of the supergroup element is given by:

\[g_{OSp}^{ - 1}  = \left( {\begin{array}{*{20}c}
   {f^{ - 1/2} w^{ - 1/2} } & { - \bar zf^{ - 1/2} w^{ - 1/2} } & {\bar \theta f^{ - 1/2} w^{ - 1/2} }  \\
   { - f^{ - 1/2} w^{1/2} } & {zf^{ - 1/2} w^{1/2} } & { - \theta f^{ - 1/2} w^{1/2} }  \\
   { - \theta f^{ - 1}  + \bar \theta f^{ - 1} } & { - z\bar \theta f^{ - 1}  + \bar z\theta f^{ - 1} } & {1 + \bar \theta \theta f^{ - 1} }  \\
\end{array}} \right)\]
\hfill (4.6)
\pagebreak

Using this representation one can form a set of supergroup one forms:

\[
\left[ J \right]_{\dot M\dot N}  = \left[ {g_{OSp}^{ - 1} dg_{OSp} } \right]_{\dot M\dot N} 
\]
\hfill (4.7)

\noindent In particular if we define the combinations of one forms as:

\[
\begin{array}{l}
 J^ +   = J_{21}\\
 J^ -   = J_{12}   \\ 
 J^3  = J_{11}  - J_{22}\\
 S^1  = J_{31}\\
 S^2  = J_{32}  \\ 
 \end{array}
\]
\hfill (4.8)

\noindent then the Cartan equations for the supergroup are [11,54]:

\[
\begin{array}{l}
 dJ^a  = \varepsilon _{    bc}^a J^b J^c  + \bar S\Gamma ^a S \\[.4cm] 
 dS = J^a \Gamma _a S \\ 
 \end{array}\]
\hfill (4.9)

\noindent Where we have introduced the target space gamma matrices:
\[
\Gamma ^a  = \left( {\begin{array}{*{20}c}
   0 & 1  \\
   { - 1} & 0  \\
\end{array}} \right),      \left( {\begin{array}{*{20}c}
   0 & 1  \\
   1 & 0  \\
\end{array}} \right),      \left( {\begin{array}{*{20}c}
   1 & 0  \\
   0 & { - 1}  \\
\end{array}} \right)
\]
\hfill (4.10)

\noindent for the signature $( - 1,1,1)$.

In the representation (4.5) the currents are given by [11]:
\begin{eqnarray*}
 J^ +   &=& f^{ - 1} e^u (dz + \theta d\theta ) \\ 
 J^ -   &=& f^{ - 1} e^{ - u} (d\bar z + \bar \theta d\bar \theta ) \\ 
 J^3 &=& du + f^{ - 1} (dz + d\bar z + \theta d\theta  + \bar \theta d\bar \theta ) \\ 
 S^1  &=& f^{ - 1/2} e^{u/2} (d\theta  - (\theta  - \bar \theta )J^ +  ) \\ 
 S^2  &=& f^{ - 1/2} e^{ - u/2} (d\bar \theta  + (\theta  - \bar \theta )J^ -  ) 
 \end{eqnarray*}
\hfill (4.11)

\noindent and we have defined u from $w = e^u $. 
\pagebreak

From these cartan generators we can obtain the target space backgrounds from the expansion:

\begin{eqnarray*}
 J_M^a dZ^M  &=& E_z^a dz + E_{\bar z}^a d\bar z + E_u^a du + E_\theta ^a d\theta  + E_{\bar \theta }^a d\bar \theta  \\ 
 S_M^\alpha  dZ^M &=& E_z^\alpha  dz + E_{\bar z}^\alpha  d\bar z + E_u^\alpha  du + E_\theta ^\alpha  d\theta  + E_{\bar \theta }^\alpha  d\bar \theta  \\ 
               &=& \Psi _z^\alpha  dz + \Psi _{\bar z}^\alpha  d\bar z + \Psi _u^\alpha  du + \Psi _\theta ^\alpha  d\theta  + \Psi _{\bar \theta }^\alpha  d\bar \theta  \\ 
 \end{eqnarray*}
\hfill (4.12)

\noindent We have used the notation $\Psi _M^\alpha   = E_M^\alpha  $
to compare with the effective action description of section II. Explicitly the background fields on $AdS^3$ are:

\[
\begin{array}{l}
  \\ 
 E_{\left( M \right)}^a  = \left( {\begin{array}{*{20}c}
   {E_Z^ +  } & {E_Z^ -  } & {E_Z^3 }  \\
   {E_{\bar Z}^ +  } & {E_{\bar Z}^ -  } & {E_{\bar Z}^3 }  \\
   {E_u^ +  } & {E_u^ -  } & {E_u^3 }  \\
\end{array}} \right) = \left( {\begin{array}{*{20}c}
   {e^u f^{ - 1} } & 0 & {f^{ - 1} }  \\
   0 & {e^{ - u} f^{ - 1} } & {f^{ - 1} }  \\
   0 & 0 & 1  \\
\end{array}} \right) \\ 
 E_{\left( \theta  \right)}^a  =
 \left( {\begin{array}{*{20}c}
   {E_\theta ^ +  } & {E_\theta ^ -}  &   { E_\theta ^3 }  \\
   {E_{\bar \theta }^ +  } & {E_{\bar \theta }^ -}&  {E_{\bar \theta }^3  }  \\
\end{array}} \right)
 = f^{ - 1}
 \left( {\begin{array}{*{20}c}
   \theta  & 0  & \theta   \\
   0 & \bar \theta &    \bar \theta    \\
\end{array}} \right) \\ 
  \\ 
\Psi _{\left( M \right)}^\alpha   = \left( \begin{array}{*{20}c}
 \Psi _Z^1 &    \Psi _Z^2  \\ 
 \Psi _{\bar Z}^1  &   \Psi _{\bar Z}^2  \\ 
 \Psi _u^1   &  \Psi _u^2  \\ 
 \end{array} \right) = (\theta  - \bar \theta )f^{ - 1/2}
 \left( \begin{array}{*{20}c}
 e^{u/2} E_z^ + &   0 \\ 
 0 & - e^{ - u/2} E_{\bar z}^ -   \\ 
 0  &     0 \\ 
 \end{array} \right) \\ 
 \Psi _{\left( \theta  \right)}^\alpha   = \left( {\begin{array}{*{20}c}
   {\Psi _\theta ^1 } & {\Psi _\theta ^2 }  \\
   {\Psi _{\bar \theta }^1 } & {\Psi _{\bar \theta }^2 }  \\
\end{array}} \right)
 = f^{ - 1/2} 
\left( {\begin{array}{*{20}c}
   {e^{u/2} } & 0  \\
   0 & {e^{ - u/2} }  \\
\end{array}} \right) \\ 
 \end{array}
\]
\hfill (4.13)

To obtain a sigma model action in the the group manifold approach one needs to form group valued currents from the Cartan one forms. To derive currents from the Cartan generators one simply has to replace $dZ^M $
with $\partial _\mu  Z^M $
 and it conventional to use the same notation for both as $J_\mu ^a  = J_M^a \partial _\mu  Z^M $. The GS action in terms of currents of the supergroup is expressed as :

\[S_{SM}  = \int {d^3 \sigma \sqrt { - g} g^{\mu \nu } J_\mu ^a } J_{a\mu }  - \frac{1}{2}\sqrt { - g}  + \int\limits_4 {WZ_4 } \]
\hfill (4.14)
\pagebreak

\noindent In this formula the Wess-Zumino four form is given by [55,56]${WZ}_4  = \bar S\Gamma _{ab} SJ^a J^b $
. On the other hand we can use the Cartan equations to relate ${WZ}_4  = d(WZ_3)= d(\varepsilon _{abc} J^a J^b J^c )$
. Because of this identity we can write the Wess-Zumino term as a three form and directly add it to the the action as:

\[S_{SM}  = \int {d^3 \sigma (\sqrt { - g} g^{\mu \nu } J_\mu ^a } J_{a\mu }  - \frac{1}{2}\sqrt { - g}  + \varepsilon _{abc} \varepsilon ^{\mu \nu \lambda } J_\mu ^a J_\nu ^b J_\lambda ^c )\]
\hfill (4.15)

\noindent Now using the $(+,-,3)$ basis for $OSp(2|1;R)$ the action becomes:

\[
S_{SM}  = \int {d^3 \sigma ( - \sqrt { - g} g^{\mu \nu } J_\mu ^ +  } J_\nu ^ -   + \frac{1}{4}\sqrt { - g} g^{\mu \nu } J_\mu ^3 J_\nu ^3  - \frac{1}{2}\sqrt { - g}  + \varepsilon ^{\mu \nu \lambda } J_\mu ^ +  J_\nu ^ -  J_\lambda ^3 )\]
\hfill (4.16)

\noindent Note that the $J_\mu ^3 $
 current can be expressed in terms of $J_\mu ^ +  $
, $J_\mu ^ -  $
 and the $u$ field so that we can write the Lagrangian as $K + WZ_3 $
with Kinetic term:

\begin{eqnarray*}
 K = \frac{1}{4}( - 2\sqrt { - g}  + \sqrt { - g} (g^{\mu \nu } J_\mu ^ +  J_\nu ^ -   - g^{\mu \nu } J_\mu ^ -  J_\nu ^ +   + g^{\mu \nu } J_\mu ^ +  J_\nu ^ +  e^{ - 2u}  + g^{\mu \nu } J_\mu ^ -  J_\nu ^ -  e^{2u}  \\ 
 \hspace{.3cm} + g^{\mu \nu } \partial _\mu  u\partial _\nu  u + g^{\mu \nu } \partial _\mu  u(J_\nu ^ +  e^{ - u}  + J_\nu ^ -  e^u ) + g^{\mu \nu } (J_\mu ^ +  e^{ - u}  + J_\mu ^ -  e^u )\partial _\nu  u)  
 \end{eqnarray*}
\hfill (4.17)

\noindent and the Wess Zumino term is given by: ${WZ}_3  = \varepsilon ^{\mu \nu \lambda } J_\mu ^ +  J_\nu ^ -  \partial _\lambda  u$
. The tranformation properties of the currents are given by [11]:
\begin{eqnarray*}
 \delta J^ +  &=& S^1 (2f^{1/2} w^{ - 1/2} \delta \theta ) + J^ +  \left( {(\theta  - \bar \theta )f^{ - 1} (\delta \theta  + \delta \bar \theta )} \right) \\ 
 \delta J^ -  &=& S^2 (2f^{1/2} w^{1/2} \delta \bar \theta ) - J^ -  \left( {(\theta  - \bar \theta )f^{ - 1} (\delta \theta  + \delta \bar \theta )} \right) \\ 
 \delta J^3  &=& \delta J^ +  w^{ - 1}  + \delta J^ -  w \\ 
 & =& S^1 (2f^{1/2} w^{ - 3/2} \delta \theta ) + S^2 (2f^{1/2} w^{3/2} \delta \bar \theta ) \\ 
&&+ J^ +  \left( {(\theta  - \bar \theta )f^{ - 1} w^{ - 1} (\delta \theta  + \delta \bar \theta )} \right) - J^ -  \left( {(\theta  - \bar \theta )f^{ - 1} w(\delta \theta  + \delta \bar \theta )} \right)  
 \end{eqnarray*}
\hfill (4.18)

\noindent So that there is alternative way of formulating $\kappa $
ivariance (4.4) directly in the supergroup sigma model. Writing out all the terms in terms of $z,\bar z,\theta ,\bar \theta $
 coordinates we have the action:
\pagebreak
\[
\hspace{-6.0cm}S_{SM} = \int d^3 \sigma \Big(- \frac{1}{2}\frac{{(\partial z + \theta \partial \theta )(\partial \bar z + \bar \theta \partial \bar \theta )}}{{(z - \bar z + \bar \theta \theta )^2 }} + \frac{1}{4}\left( {\partial u} \right)^2\]
\[
\hspace{3.5cm}+\frac{1}{2}\frac{{\partial u(\partial z + \theta \partial \theta  + \partial \bar z + \bar \theta \partial \bar \theta )}}{{(z - \bar z + \bar \theta \theta )}} + \frac{1}{4}\frac{{(\partial z + \theta \partial \theta )^2 }}{{(z - \bar z + \bar \theta \theta )^2 }} + \frac{1}{4}\frac{{(\partial \bar z + \bar \theta \partial \bar \theta )^2 }}{{(z - \bar z + \bar \theta \theta )^2 }}\]
\[
\hspace{9.5cm}- \frac{1}{2}\sqrt { - g}  + \frac{{(dz + \theta d\theta )(d\bar z + \bar \theta d\bar \theta )du}}{{(z - \bar z + \bar \theta \theta )^2 }} \Big)\]
\hfill (4.19)

\noindent where we have used the notation $\partial a\partial b = da * db = \sqrt { - g} g^{ij} \partial _i a\partial _j b$
.

\subsection*{Background Fields}

From the WV action (4.19)  the target space metric can be read off directly by expanding to lowest order in $\theta $
. Comparison with (4.19) yields:

\begin{eqnarray*}
 ds_3^2 &=& G_{MN} dX^M dX^N \\ 
&=&  - \frac{1}{2}\frac{{dzd\bar z}}{{(z - \bar z)^2 }} + \frac{1}{4}du^2  + \frac{1}{2}\frac{{du(dz + d\bar z)}}{{(z - \bar z)^2 }} + \frac{1}{4}\frac{{\left( {dz} \right)^2 }}{{(z - \bar z)^2 }} + \frac{1}{4}\frac{{\left( {d\bar z} \right)^2 }}{{(z - \bar z)^2 }} \\ 
&=&  - \frac{{dzd\bar z}}{{(z - \bar z)^2 }} + \frac{1}{4}(du + \frac{{dz + d\bar z}}{{z - \bar z}})^2  \\ 
 \end{eqnarray*}
\hfill (4.20)

\noindent This is the same target space background found from the effective action consideratons of section II so that the background solves the target space equations of motion (2.12). Being three dimensional however the gravitational background is not associated to any target space states, there are no gravitons in three dimensions. 

The bosonic $T$ field and or its superpartner lead to a single degree of freedom. It's equation can be obtained by noting that the stress energy tensor is quadratic in the currents and is written:

\[
\begin{array}{l}
 T_{\mu \nu }  =  - J_\mu ^ +  J_\nu ^ -   - J_\mu ^ -  J_\nu ^ +   + J_\mu ^ +  J_\nu ^ +  e^{ - 2u}  + J_\mu ^ -  J_\nu ^ -  e^{2u}  \\ 
  + \partial _\mu  u\partial _\nu  u + \partial _\mu  u(J_\nu ^ +  e^{ - u}  + J_\nu ^ -  e^u ) + (J_\mu ^ +  e^{ - u}  + J_\mu ^ -  e^u )\partial _\nu  u - \frac{1}{2}g_{\mu \nu } K \\ 
 \end{array}
\]
\hfill (4.21)

\noindent The zero mode of this stress energy tensor is denoted by $L_0 $
 and provides a link between the sigma model and effective action approaches [57]. In terms of the target space metric of the group manifold the bosonic component of the $L_0 $
 operator is denoted $L_0^B $
 and it is given by the differential operator [57];

\[
L_0^B  =  - \frac{1}{{2\sqrt { - G} }}\frac{\partial }{{\partial X^M }}(\sqrt { - G} G^{MN} \frac{\partial }{{\partial X^N }})
\]
\hfill (4.22)

\noindent As the sigma model does not contain a Einstein-Hilbert  term the stress tensor should be zero on shell as an equation of motion found from varying $g$. In particular the zero mode of the stress tensor should be zero and we obtain the mass shell condition $(L_0^B )T(X) = 0$
which in the parametrization (4.5) becomes:

\[
(L_0^B )T(X) = 4(z - \bar z)^2 \frac{\partial }{{\partial z}}\frac{\partial }{{\partial \bar z}}T + 2\frac{{\partial ^2 }}{{\partial u^2 }}T - 4(z - \bar z)(\frac{\partial }{{\partial z}} + \frac{\partial }{{\partial \bar z}})\partial _u T = 0
\]
\hfill (4.23)

\noindent This expression is In agreement with the equation of motion for $T$ derived from the effective action (2.12) in the background (2.15).  

The process of double dimensional reduction on the coordinate $u$
to a IIA string theory can also be see from this equation. The $L_0 $
operator becomes $L_{0IIA}  = \frac{1}{{e^{ - 2\Phi } \sqrt { - G^{(2)} } }}\partial _I (e^{ - 2\Phi } \sqrt { - G^{(2)} } G^{(2)IJ} \partial _J )$
so that $ds_2^2  = G_{IJ}^{(2)} dX^I dX^J  = \frac{{dZd\bar Z}}{{(Z - \bar Z)^2 }}$ ans $\Phi  = \Phi _0 $ in agreement with section III.

Before discussing amplitudes it will be useful to define Pioncaire coordinates through:

\[g_{SL2}  = \left( {\begin{array}{*{20}c}
   1 & {\bar \gamma }  \\
   0 & 1  \\
\end{array}} \right)\left( {\begin{array}{*{20}c}
   {e^\varphi  } & 0  \\
   0 & {e^{ - \varphi } }  \\
\end{array}} \right)\left( {\begin{array}{*{20}c}
   1 & 0   \\
   \gamma & 1  \\
\end{array}} \right) = \left( {\begin{array}{*{20}c}
   {e^{ - \varphi }  + e^\varphi  \gamma \bar \gamma } & {e^\varphi  \bar \gamma }  \\
   {e^\varphi  \gamma } & {e^\varphi  }  \\
\end{array}} \right)\]
\hfill (4.24)

\noindent In these coordinates the Target space metric is given by :

\[ds_3^2  = G_{MN} dX^M dX^N  = (d\varphi ^2  + e^{2\varphi } d\gamma d\bar \gamma )\]
\hfill (4.25)

\noindent and the relation to the previous parametrization is $z = \gamma ^{ - 1} e^{ - 2\varphi }  + \bar \gamma ,\bar z = \bar \gamma ,e^u = \gamma$. 
\pagebreak

The equation for the $T$ field is given in Poincare coordinates as:

\[
(\pi _\varphi ^2  + 4e^{2\varphi } \pi _\gamma  \pi _{\bar \gamma }  - 4j(j - 1))T(X) = (W^2 \frac{{\partial ^2 }}{{\partial W^2 }} - W\frac{\partial }{{\partial W}} + 4W^2 \frac{\partial }{{\partial \gamma }}\frac{\partial }{{\partial \bar \gamma }} - 4j(j - 1))T(X) = 0\]
\hfill (4.26)

\noindent and we have defined $W = e^\varphi  $
. This equation has the solution of the form: 

\[
T(X) = e^{i\nu _ +  \gamma } e^{i\nu _ -  \bar \gamma } e^{ - \varphi } K_{2j + 1} (2\sqrt {\nu _ +  \nu _ -  } e^{ - \varphi } )
\]
\hfill (4.27)

\noindent Here $K_{2j+1}$ is the modified Bessel function. In this expression we have allowed the field $T$ to go off shell by an amount proportion to $j(j - 1)$
the eigenvalue of the Casimir operator on $SL(2,R)$. 

\subsection*{Amplitudes}

Now returning to the action we can define a quantum amplitude for the supermembrane on $AdS^3$ from the path integral:

\begin{eqnarray*}
\lefteqn{ A(j_1 ,\nu _{ + 1} ,\nu _{ - 1} , \ldots ,j_n ,\nu _{ + n} ,\nu _{ - n} ) =  }\\ 
	& & \int {DgDZD\bar Z} DuD\theta D\bar \theta e^{iS} \int {d^3 \sigma _1 }  \ldots \int {d^3 \sigma _n V_1 } (\sigma _1 ,j_1 ,\nu _{ + 1} ,\nu _{ - 1} ) \ldots V_n (\sigma _n ,j_n ,\nu _{ + n} ,\nu _{ - n} ) \\ 
 \end{eqnarray*}
\hfill (4.28)

\noindent The vertex operators are defined by in the the supergroup through:

\[
V(\sigma ,j,\nu _ +  ,\nu _ -  ) = \Phi _j^{\nu _ +  \nu _ -  } (Z(\sigma ),\bar Z(\sigma ),u(\sigma ),\theta (\sigma ),\bar \theta (\sigma )) = \left\langle {j,\nu _ +  |g_{Osp} (Z,\bar Z,u,\theta ,\bar \theta )|j,\nu _ -  } \right\rangle \]
\hfill (4.29)

\noindent In this expression $j,\nu _ +  ,\nu _ -  $
are quantum numbers that play the role of energy and momentum on $AdS^3$. 

To proceed further we need an explicit expression for the vertex operators for a supermembrane on $AdS^3$. For string theory on $AdS^3$ the group manifold transformation properties and representation theory can be used to construct the vertex operators [58]. Similar considerations apply here. The group manifold approach to vertex operators is most conveniently expressed in terms of the Poincare corrdinate system. First consider the bosonic piece of the group element given by $g_{SL2} $. In this representation one can form the combination:

\[F(x,\bar x,g_{SL2} ) = (1, - x)g_{SL2} \left( \begin{array}{l}
 1 \\ 
  - \bar x \\ 
 \end{array} \right)\]
\hfill (4.30)

\noindent for two real parameters $x$
 and $\bar x$. Primary fields can then be contructed using the integral representation:

\[
\Phi _j^{\nu _ +  \nu _ -  }  = \int {dxd\bar x} e^{i\nu _ +  x + i\nu _ -  \bar x} |F(x,\bar x,g_{SL2} )|^{ - 2j} \]
\hfill (4.31)

\noindent In terms Poincare coordinates $z = \gamma ^{ - 1} e^{ - 2\varphi }  + \bar \gamma ,\bar z = \bar \gamma ,e^u = \gamma$ we have $F(x,\bar x,g_{SL2} ) = (\bar \gamma  - \bar x)(\gamma  - x)e^\varphi   + e^{ - \varphi } $
 so that:

\[
\Phi _j^{\nu _ +  \nu _ -  }  = \int {dxd\bar x} e^{i\lambda x + i\mu \bar x} |(\bar \gamma  - \bar x)(\gamma  - x)e^\varphi   + e^{ - \varphi } |^{ - 2j}
 \]
\hfill (4.31)

\noindent After performing the integration one obtains the expression: 

\[
\left\langle {j,\lambda } \right.|g_{SL2} (\varphi ,\gamma ,\bar \gamma )|j,\left. \mu  \right\rangle  = \Phi _j^{\lambda \mu } (g) = e^{i\lambda \gamma } e^{i\mu \bar \gamma } e^{ - \varphi } K_{2j + 1} (2\sqrt {\lambda \mu } e^{ - \varphi } )
\]
\hfill (4.32)

Proceeding in a similar manner to for the supergroup $OSp(2|1;R)$ we introduce a pair of Grassman parmeters $\xi ,\bar \xi $
and form the expression:

\[F(x,\bar x,\xi ,\bar \xi ,g_{OSp} ) = (1, - x,\xi )g_{OSp} \left( \begin{array}{l}
 1 \\ 
  - \bar x \\ 
 {\bar \xi } \\ 
 \end{array} \right)\]
\hfill (4.33)

\noindent The primary operators can be formed from the integral:

\[
\Phi _j^{\nu _ +  \nu _ -  }  = \int {dxd\bar x} d\xi d\bar \xi e^{i\nu _ +  x + i\nu _ -  \bar x} \frac{1}{{ - 2j + 1}}|F(x,\bar x,\xi ,\bar \xi ,g_{OSp} )|^{ - 2j + 1} 
\]
\hfill (4.34)

\noindent and because of the Grassman integration only the coefficient of $\xi \bar \xi $
in the expansion of the integrand gives a nonzero result. Performing this expansion and using $f=z-\bar z + \bar \theta \theta=e^{-2\varphi}\gamma^{-1}+\bar \theta \theta$ we find a correspoding term for the $T$ vertex given by:

\begin{eqnarray*}
 \Phi _j^{\nu _ +  \nu _ -  } &=& \int {dxd\bar x} d\xi d\bar \xi e^{i\nu _ +  x + i\nu _ -  \bar x} |F(x,\bar x,g_{SL2} )|^{ - 2j} (1 + \bar \theta \theta e^{2\varphi } \gamma )\xi \bar \xi  \\ 
 &=& \int {dxd\bar x} e^{i\nu _ +  x + i\nu _ -  \bar x} |F(x,\bar x,g_{SL2} )|^{ - 2j} (1 + \bar \theta \theta e^{2\varphi } \gamma ) \\ 
  \\ 
 \end{eqnarray*}
\hfill (4.35)

\noindent Now using the result (4.32) we have the expression for the membrane vertex operator on $AdS^3$:

\[
V(j,\nu _ +  ,\nu _ -  ) = \left\langle {j,\nu _ +  } \right.|g_{OSp} (\varphi ,\gamma ,\bar \gamma ,\theta ,\bar \theta )|j,\left. {\nu _ -  } \right\rangle  = e^{i\nu _ +  x + i\nu _ -  \bar x} K_{2j + 1} (2\sqrt {\nu _ +  \nu _ -  } e^{ - \varphi } )(1 + \bar \theta \theta e^{2\varphi } \gamma )
\]
\hfill (4.36)

The relation of the form of the vertex operator and the solution to the field equation is in accordance with mapping of operators to states in the theory. The main distinction between the supermembrane amplitude and string theory amplitudes on $AdS^3$ is the larger world volume symmetries of the theory, the integration over $d^3 \sigma $
instead of $d^2 \sigma $
and the specification of three metric of the world volume in (4.28). In a more general context covariant and lightcone supermembrane vertex operators can be used to define the amplitudes (4.28) for the supermembrane in  background fields [50,59]. Again the difference between the supermembrane on $AdS^3 $
and for example, the M2 supermembrane on $AdS^4  \times S^7 $
 [55,56] is the lack of transverse modes on $AdS^3$, which indicates that the path integral for the supermembrane on $AdS^3 $
is potentially finite.
\pagebreak

\section*{V Alternative World Volume actions and 3d gravity}

In the review of two dimensional noncritical string theory in section III the Green Schwarz superstring on $AdS^2$ was considered, as well as alternative NSR spinning string and the WS 2d gravity actions. In particular,  the original M-atrix formulation for two dimensional string theory arose from the random lattice approach to the WS 2d gravity action. Thus for three dimensional M-theory it is also of interest to investigate the existence of alternative world volume actions of the NSR and WV 3d gravity type.

\subsection*{WV NSR spinning membrane}

First consider the  WV NSR type action given by Howe and Tucker [60]:

\begin{eqnarray*}
\lefteqn{S_{NSR}  =  \int {d^3 } \sigma \{ \det (e)\{ \frac{1}{2}g^{\mu \nu } j_\mu ^M j_\nu ^M + \bar \psi ^M \gamma ^\mu  \nabla _\mu  \psi ^M}\hspace{2.5cm}\\
		& &+ \frac{i}{2}j_\mu ^M \bar \psi ^M \gamma ^\mu  \gamma ^\nu  \chi _\nu + \frac{1}{8}\bar \chi _\mu  \gamma ^\nu  \gamma ^\mu  \psi ^M \bar \chi _\nu  \psi ^M \}\\
		& &+ i\frac{1}{4}\varepsilon ^{\mu \nu \rho } \bar \chi _\mu  \gamma _\nu  \chi _\rho - \det (e)\frac{1}{4}i\bar \psi ^M \psi ^M  + \frac{1}{2}\det (e)\} \\
\end{eqnarray*}
\hfill (5.1)

\noindent with $j_\mu ^M  = E_N^M \partial _\mu  X^N $
. This was the original action used to study the spinning membrane. The WV spinors have vectors indices so that Target Space supersymmetry is more difficult to address than in the GS form (4.1) The theory has also a difficulty with the absence of a local world volume supersymmetry [61,62] and thus is not nearly as well studied as the GS type supermembrane. A modifcation has been condsidered to the spinning membrane that included three dimensional Weyl invariance [63]. However without the elimination of degrees of freedom due to a local fermionic symmetry on the world volume it is difficult to obtain a target space supersymmetric theory, as bosonic degrees of freedom are already eliminated due to the  local world volume reparametrization invariance.

The action (5.1)  is closely related to the WS NSR action (3.7) discussed in section III. The first two lines in are familiar. They represent a minimally coupled scalar multiplet coupled to to a WV metric and Rarita Schwinger field. The bosonic part of the action is identical with (4.1) so that the difference with the GS Membrane is in the fermionic sector. The third line in (5.1) has a WV cosmological constant term but no kinetic term for the metric or Rarita Schwinger field. The absence of a Rarita-Schwinger term also occurs in the WS NSR action (3.7) and the WS GS form of the action (3.1) for string theory . However in the string WS case the absence of these terms was because the 2d Einstein-Hilbert term is an exact divergence and because the WS Rarita-Schwinger action vanishes identically in 2d. For a WV action the 3d Einstein-Hilbert term is not an exact divergence, the 3d Rarita-Schwinger action does not vanish and this leads us directly to the WV 3d gravity formulation of M-theory that we now discuss.

\subsection*{WV 3d gravity approach}

The WV 3d gravity approach is defined by replacing  the last line of the above NSR WV action with a full 3d gravity action with all kinetic terms intact to obtain:

\begin{eqnarray*}
\lefteqn{S_{3dgrav}  =   \int {d^3 } \sigma\{\det (e)\{ \frac{1}{2}g^{\mu \nu } G_{\bar M\bar N} \partial _m X^{\bar M} \partial _n X^{\bar N}  + \bar \psi ^{\bar M} \gamma ^m \nabla _m \psi ^{\bar M}  }\hspace{2.5cm}\\ 
&&+ \frac{i}{2}E_{\bar M}^{\bar N} \partial _\nu  X^{\bar M} \bar \psi ^{\bar N} \gamma ^\mu  \gamma ^\nu  \chi _\mu   + \frac{1}{8}\bar \chi _\mu  \gamma ^\nu  \gamma ^\mu  \psi ^{\bar M} \bar \chi _\nu  \psi ^{\bar M} \}  \\ 
&&+ \det (e)R - 2\lambda \det (e) +\varepsilon ^{\mu \nu \rho } \bar \chi _\mu  D_\nu  \chi _\rho   + i\sqrt{- \lambda}  \frac{1}{2}\varepsilon ^{\mu \nu \rho } \bar \chi _\mu  \gamma _\nu  \chi _\rho  \}  
 \end{eqnarray*}
\hfill (5.2)

\noindent This form of the action was also considerered by Howe and Tucker in [60]. A related action was discussed by Kogan [64] with string degrees of freedom living on a 2d boundary, as well as a M(atrix) Chern-Simons description of Livine and Smolin in [65].

The most important point about the action (5.2) versus (5.1) or (4.1) is that when one includes a kinetic term for the WV metric one effectively changes the Target space dimension. The formula relating the Target space dimension from the kinetic field count is given by:

\[D = N_L  + N_S  = (6 - 3) + N_S  = (9 - 3 - 3) + N_S  = 3 + N_S \]
\hfill (5.3)

\noindent This formula requires some explanation. Here $N_L$ is the number of Liouville fields which is the number of component of the spatial components of the world volume metric (3). $N_L$ is also given by the number of WV metric components (6) minus the number of reparametrization symmetries (3). Still another way of counting the Liouville fields is to use the dreibein components (9) minus the number of WV Local Lorentz symetries (3) minus the number of reparametrization symmetries (3). In any case the final number of target space dimensions is $N_L$ plus $N_S$ the number of scalars. Formula (5.3) is WV the analog of (3.9) used for 2d gravity WS theories. 
\pagebreak

Formula (5.3) means that for $D=3$, which is the subject of this paper, we must have $N_S  = 0$
 in the WV 3d gravity approach. In this case the WV 3d gravity action dramatically simplifies to:

\[
S_{3dgrav}  = \int {d^3 \sigma \det (e)R - 2\lambda \det (e) +\varepsilon ^{\mu \nu \rho } \bar \chi _\mu  D_\nu  \chi _\rho   + i\sqrt{- \lambda}  \frac{1}{2}\varepsilon ^{\mu \nu \rho } \bar \chi _\mu  \gamma _\nu  \chi _\rho  } 
\]
\hfill (5.4)

\noindent which is the action of pure 3d gravity with cosmological constant.

\subsection*{Target Space Metric}

To go beyond the counting argument of Formula  (5.3) it is necessary to  identify the actual target space background described by the action (5.4). Indeed the most difficult aspect of the WV 3d gravity formulation of microscopic M-theory is its target space interpretation. To identify the target space it is sufficient to study the bosonic contribution to the action, One way to approach the  target space interpretation of (5.4) is to go to a canonical gauge in terms of three Liouville type fields $\phi _1 ,\phi _2 ,\phi _3 $
 and parametrize the WV metric as :

\[g_{\mu \nu }  = \left( {\begin{array}{*{20}c}
   h & 0  \\
   0 & { - 1}  \\
\end{array}} \right) = \left( {\begin{array}{*{20}c}
   {e^{\phi _1  - \phi _2 } } & {e^{\phi _1  - \phi _2 } \phi _3 } & 0  \\
   {e^{\phi _1  - \phi _2 } \phi _3 } & {e^{\phi _1  + \phi _2 }  + e^{\phi _1  - \phi _2 } \phi _3^2 } & 0  \\
   0 & 0 & { - 1}  \\
\end{array}} \right) = \left( {\begin{array}{*{20}c}
   {Vm_2^{ - 1} } & {Vm_1 m_2^{ - 1} } & 0  \\
   {Vm_1 m_2^{ - 1} } & {Vm_2^{ - 1} (m_1^2  + m_2^2 )} & 0  \\
   0 & 0 & { - 1}  \\
\end{array}} \right)\]
\hfill (5.5)

\noindent This formula is not to be confused with the Target space metric but represents the WV analog of (3.8). In this parametrization $V = e^{\phi _1 } $
describes the size of the world volume and $m_1  = \phi _3 ,m_2  = e^{\phi _2 } $
 describe its shape or anisotropy. This form is special case of the general parametrization invovlving the Lapse and shift functions $N,N^i $
:

\[g_{\mu \nu }  = \left( {\begin{array}{*{20}c}
   {h_{ij} } & {h_{ik} N^k }  \\
   {h_{ik} N^k } & { - N^2  + h_{ij} N^i N^j }  \\
\end{array}} \right);\hspace{1.0cm} h_{ij}  = V\frac{1}{{m_2 }}\left( {\begin{array}{*{20}c}
   1 & {m_1 }  \\
   {m_1 } & {m_1^2  + m_2^2 }  \\
\end{array}} \right)\]
\hfill (5.6)

\noindent with $N^i  = 0,N = 1$ describing the canonical gauge. The analog of the conformal gauge is given by $N^i=0,N=\sqrt{\det h}$. Note $h$ parametrizes $R\times SL(2,R)/SO(2)$ through:
\[h_{R\times SL2/U1}  = V\left( {\begin{array}{*{20}c}
   1 & 0  \\
   {m_1 } & 0  \\
\end{array}} \right)\left( {\begin{array}{*{20}c}
   {m_2^{ - 1} } & 0  \\
   0 & {m_2 }  \\
\end{array}} \right)\left( {\begin{array}{*{20}c}
   1 & {m_1 }  \\
   0 & 0  \\
\end{array}} \right)\]
\hfill (5.7)

\noindent a  form similar to the product representation of $AdS^3 $
(4.24) in terms of Poincare coordinates.

The bosonic contribution to the WV action in the canonical gauge is given by:
\begin{eqnarray*}
 S_B &=& \int {d^3 \sigma \sqrt { - g} } (R - 2\lambda ) = \int {d^3 \sigma } \frac{1}{{4\sqrt {h_{11} h_{22}  - h_{12}^2 } }}(\partial _3 h_{11} \partial _3 h_{22}  - \partial _3 h_{12} \partial _3 h_{12} ) - 2\sqrt h \lambda  + \sqrt h R_h  \\ 
 &=& \int {d^3 } \sigma \frac{1}{4}e^{\phi _1 } ( - \partial _3 \phi _1 \partial _3 \phi _1  + \partial _3 \phi _2 \partial _3 \phi _2  + e^{ - 2\phi _2 } \partial _3 \phi _3 \partial _3 \phi _3 ) - 2e^{\phi _1 } \lambda  + \sqrt h R_h  \\ 
&=& \int {d^3 } \sigma \frac{1}{4}V( - V^{ - 2} \partial _3 V\partial _3 V + m_2^{ - 2} \partial _3 m_2 \partial _3 m_2  + m_2^{ - 2} \partial _3 m_1 \partial _3 m_1 ) - 2V\lambda  + \sqrt h R_h   
 \end{eqnarray*}
\hfill (5.8)

\noindent where $R_h $
is the curvature associated with the two metric $h$
so that the last term in (5.8) is a local divergence. Written in terms of the $V,m_1 ,m_2 $
 variables the Kinetic term in (5.8) describes a Target space $R \times SL(2,R)/SO(2)$
 and Target space metric

\begin{eqnarray*}
 ds_3^2 &=& G_{MN} dX^M dX^N = \sqrt h (h^{ik} h^{jl}  + h^{il} h^{jk}  - 2h^{ij} h^{kl} )dh_{ij} dh_{kl}  \\ 
 &=& e^{\phi _1 } ( - d\phi _1^2  + d\phi _2^2  + e^{ - 2\phi _2 } d\phi _3^2 ) = V( - \frac{{dV^2 }}{{V^2 }} + \frac{{dm_1^2  + dm_2^2 }}{{m_2^2 }})  
 \end{eqnarray*}
\hfill (5.9)

\noindent The Target space interpretation for the kinetic term of (5.8) goes back to the original DeWitt analysis for 4d gravity [66]. It is also implicitly used in recent work on the appearance of Coset spaces in anisotropic cosmological models. See for example work on quantum cosmology and  Cosmological Billiards [67,68]. 

Reintroducing the lapse function $N$
, the Target space background metric and $T$ field are $ds_3^2  = \frac{{e^\phi  }}{N}( - d\phi ^2  + dx^2  + e^{ - 2x} dm_1^2 )$
and $T =  - 2\lambda Ne^\phi  $
 where we have defined $V = e^\phi  $
and $m_2  = e^x $
. Applying the DDR process to the coordinate $m_1 $
, and choosing the  $N = e^\phi  $ gauge leads to the 2D backgrounds $ds_2^2  =  - d\phi ^2  + dx^2 $
, $T =  - 2\lambda e^{2\phi } $
and $\Phi  = x$
. Also the metric $ds_3^2 $
 in this form is one of the Thurtston three geometries $R\times H^2  \sim R\times EAdS^2 $
and can be expressed as a solution to a target space theory with $F_{(2)} $
and $F_{(3)} $
fields [43] or as a squashed version of $AdS^3 $
 [69].

Equation (5.9) indicates that the WV 3d gravity approach to M-theory in $D=3$ comes with its own background, like in the WS 2dgravity approach  to string theory in $D=2$,. How could one possibly probe other backgrounds? In the 2d gravity Liouville approach to string theory this was also an issue, as the Liouville action is strictly written in a linear dilaton background. Yet in that case the presence of other discrete states in the theory, a remnant of a target space graviton at discrete momentum, allowed one to probe other backgrounds with the Liouville theory, namely a $D=2$ black hole [70]. Thus it may turn out that a remnant of a Target space graviton in $D=3$ also exists at discrete momentum, (no gravitons exist at arbitrary continuous momentum in $D=2$ or $D=3$). Perturbations involving discrete states may  allow one to move away from the $R\times SL(2,R)/SO(2)$ target space described by (5.9).

\subsection*{State Analysis, T Field equation, Liouville and CS WV 3d gravity actions}

The next topic we wish to discuss is the onshell spectrum in the theory. There are many approaches to the state analysis of pure WV 3d gravity. Perhaps the simplest method is the gauge fixed Wheeler deWitt approach of Martinec [71]. In this case there is one state $T$ describing empty WV and no oscillators. The bosonic component of the field equation for $T$ is then given by the WV Wheeler-deWitt (WdW) equation in the $X = (m_1 ,m_2 ,V)$
 parametrization (5.5) :

\[(V^{ - 1/2} \pi _V V^{3/2} \pi _V  - \frac{{m_2^2 }}{V}\pi _1^2  - \frac{{m_2^2 }}{V}\pi _2^2  - 2\lambda V + \sqrt h R^{(2)} )T(X) = 0\]
\hfill (5.10)

\noindent Like in (4.23) the zero mode piece of this constraint plays the role of the $L_0  = 0$
physical state condition and can be expressed as a differential operator:

\[( - V\frac{{\partial ^2 }}{{\partial V^2 }} - \frac{3}{2}V\frac{\partial }{{\partial V}} + \frac{1}{V}m_2^2 (\frac{{\partial ^2 }}{{\partial m_1^2 }} + \frac{{\partial ^2 }}{{\partial m_2^2 }}) - 2\lambda V + 8\pi (1 - g_e ))T(X) = 0
\]
\hfill (5.11)

\noindent Here $g_e $
is the genus associated with the spatial two surface assuming a decomposition $WV = R \times \Sigma $
. The equation represents the Laplacian for a scalar field in the background (5.9). The single state description is consistent with the effective action analysis of section II.

For the simplest case of $\Sigma  = T^2 $
we have $g_e  = 1$
and the spatial curvature term is absent. The equation has a solution of the form [69]:

\[
T(X) = \sum\limits_{p,n} {a_{p,n} V^{1/2} J_{ \pm i\sqrt {2p} } (i\sqrt { - \lambda } V)e_{p,n} (m_1 ,m_2 )}  + c.c.
\]
\hfill (5.12)

\noindent with $J$ the Bessel function and $e_{p,n} (m_1 ,m_2 )$
solves the equation $\Delta _{H^2 } e_{p,n}  = (p^2  + \frac{1}{8})e_{p,b} $
which is the Laplacian on the hyperblic space $H^2 $
. The $e_{p,n} (m_1 ,m_2 )$
 are given as linear combinations of $u_p^{(n)} (m_1 ,m_2 ) = m_2^{1/2} K_{ip} (2\pi |n|m_2 )e^{2\pi inm_1 } $
and $h_p^ \pm  (m_1 ,m_2 ) = m_2^{1/2} e^{ \mp ip\ln m_2 } $
[44]. More general representations of the states can be obtained in an arbitrary gauge by solving the constraint algebra of 3d supergravity [72].

It is interesting to note that conformal quantum mechanics plays a role in WV 3d gravity [67] just as it does for WS 2d string theory [11,36].  Defining the combinations: 
\begin{eqnarray*}
 E_ -   &=& \frac{1}{2}(V^{1/2} \pi _V V^{1/2} \pi _V  - \frac{1}{V}m_2^2 (\pi _1^2  + \pi _2^2 )) - 2\lambda V \\ 
 E_ +  &=& 2V \\ 
 D_0  &=& V\pi _V   
 \end{eqnarray*}
\hfill (5.13)

\noindent these operators generate the SO(2,1) algebra $\{ E_ +  ,E_ -  \}  = 2D_0 ,\{ D_0 ,E_ \pm  \}  =  \pm E_ \pm  $
. The relation to conformal quantum mechanics is clearer with the change of variables $q = 2V^{1/2} ,p = V^{1/2} \pi _V $
where the generators take the form [67,73]:
\begin{eqnarray*}
 E_ -  & =& \frac{1}{2}(p^2  + \frac{{g^2 }}{{q^2 }}) - 2\lambda q^2  \\ 
 E_ +  & =& \frac{1}{2}q^2  \\ 
 D_0  &=& \frac{1}{2}qp 
 \end{eqnarray*}
\hfill (5.14)

\noindent and $g = 4\Delta  =  - 4m_2^2 (\pi _1^2  + \pi _2^2 )$
. More details of the representation of the anti-deSitter algebra from 3d gravity can be found in [74].

Now returning to the action with fermions we find two other forms of the world volume action can be derived from (5.4 that are more amenable to calculation. Given the action (5.4) in the gauge (5.5) we can write the WV 3dgrav action in a WV superliouville form:
\begin{eqnarray*}
\lefteqn{ S_L = \int {d^3 } \sigma \{ \frac{1}{4}e^{\phi _1 } ( - \partial _3 \phi _1 \partial _3 \phi _1  + \partial _3 \phi _2 \partial _3 \phi _2  + e^{ - 2\phi _2 } \partial _3 \phi _3 \partial _3 \phi _3 ) + \sqrt h R_h}\hspace{2.5cm}  \\ 
&&  + \varepsilon ^{ijk} \bar \chi _i D_j \chi _k  + \frac{i}{2}\sqrt { - \lambda } e^{\phi _1 /2} \tilde e_i^a \varepsilon ^{ijk} \bar \chi _j \gamma_a \chi _k  - 2\lambda e^{\phi _1 } \}  \\ 
 \end{eqnarray*}
\hfill (5.15)

\noindent where we have defined:
\[
\tilde e_i^a (\phi _2 ,\phi _3 ) = e^{ - \phi _2 /2} \left( {\begin{array}{*{20}c}
   1 & 0 & 0  \\
   {\phi _3 } & {e^{\phi _2 } } & 0  \\
   0 & 0 & 1  \\
\end{array}} \right)
\]
\hfill (5.16)

\noindent This form (5.15) is closely analogous to the WS superliouville action given by (3.4). Thus we see from (5.15) that in a sense the microscopic M-theory WV 3dgavity action for $D=3$ is pure Liouville. 

The final form of the WV 3d gravity action is obtained by expressing the theory  in a $Osp(2|1;R\times SO(2,1)$ Chern-Simons formulation [75,76,77]. In this form the action is written

\begin{eqnarray*}
 S_{CS}^ +  & =& \int {d^3 \sigma \varepsilon ^{\mu \nu \rho } (A_\mu ^{ + a} } \partial _\nu  A_{\rho a}^ +   + \frac{1}{3}\varepsilon _{abc} A_\mu ^{ + a} A_\nu ^{ + b} A_\rho ^{ + c}  + \bar \chi _\mu ^ +  \partial _\nu  \chi _\rho ^ +   + \frac{i}{2}A_\mu ^{ + a} \bar \chi _\nu ^ +  \gamma _a \chi _\rho ^ +  ) \\ 
 S_{CS}^ -  & =& \int {d^3 \sigma \varepsilon ^{\mu \nu \rho } (A_\mu ^{ - a} } \partial _\nu  A_{\rho a}^ -   + \frac{1}{3}\varepsilon _{abc} A_\mu ^{ - a} A_\nu ^{ - b} A_\rho ^{ - c} ) \\ 
 \end{eqnarray*}
\hfill (5.17)

\noindent with the Chern-Simons one form related to the dreibein and spin connection through $ A_\mu ^{ \pm a}  = \omega _\mu ^a  \pm \mu e_\mu ^a$ and $ 
 \chi _\mu ^{ + \alpha } = \sqrt {2\mu } \chi _\mu ^\alpha $

\noindent The WV 3dgrav action is then given by the difference:
\begin{eqnarray*}
 S_{grav}  &=& \frac{1}{2}(S_{CS}^ +   - S_{CS}^ -  ) = \int {e(d\omega  + \omega ^2 ) + \bar \chi D\chi  + 2\mu ^2 e^3  + i\frac{1}{2}\mu e^a \bar \chi \gamma _a \chi}  \\ 
 &=& \int {\varepsilon ^{\mu \nu \rho } e_\mu ^a (\partial _\nu  \omega _{a\rho }  - \partial _\rho  \omega _{a\nu }  + \varepsilon _{abc} \omega _\nu ^b \omega _\rho ^c ) - 2\lambda \varepsilon ^{\mu \nu \rho } \varepsilon _{abc} e_\mu ^a e_\nu ^b e_\rho ^c }  \\ 
&& +  \varepsilon ^{\mu \nu \rho } \bar \chi _\mu  (\partial _\nu   + \frac{1}{2}\omega _\nu ^a \gamma _a )\chi _\rho   + i\frac{1}{2}\sqrt { - \lambda } \varepsilon ^{\mu \nu \rho } \bar \chi _\mu  e_\nu ^a \gamma_a \chi _\rho   \\ 
 \end{eqnarray*}
\hfill (5.18)

\noindent and we have used the notation of differential forms in the first line of (5.18).

It is evident that the double dimensional reduction process applied to the WV action (5.18) does not yield the GS string of (3.1) but instead the 2d WS action [78]:

\begin{eqnarray*}
 S_{DDR}  &=& \int {d^2 \sigma \sqrt { - g} } X(R - 2\lambda ) + i\frac{1}{2}\mu X\varepsilon ^{ij} \bar \chi _i \gamma _3 \chi _j  + \bar \psi \varepsilon ^{ij} D_i \chi _j  \\ 
 & =& \int {d^2 \sigma \sqrt { - \hat g} } ( - 2\partial X\partial \phi  + X\hat R - 2\lambda Xe^\phi  ) + i\frac{1}{2}\mu X\varepsilon ^{ij} \bar \chi _i \gamma _3 \chi _j  + \bar \psi \varepsilon ^{ij} D_i \chi _j  \\ 
 \end{eqnarray*}
\hfill (5.19)

\noindent with $X=\sqrt{g_{33}}$ and $\psi=\chi_3$. After applying the conformal gauge as in section III this is recognized as a variant of the WS gravity approach with WS fields $(\phi ,X,\zeta ,\psi )$
. The reduced 2d action (5.19) was also related to superconformal quantum mechanics in [78].

Thus if one attempts a DDR with the action (5.15) or (5.17) or the field equation (5.11) one obtains a WS theory in two target dimensions of the WS 2d gravity type or a 2d sigma model with two target dimensions. For world volumes of the form $WV = [0,1] \times \Sigma $
 the reduction takes a special form and the WS theory is related to the superliouville expression of 2d induced gravity [53]. Again one does not obtain a 2d action of the Green Schwarz type superstring and this differs from the supermembrane action discussed in the previous section.

\subsection*{Amplitudes}

The Chern-Simons formulation with gauge group $OSp(2|1;R)\times SO(2,1)$ shows that the path integral using the  WV action (5.19) interpreted as $D=3$ Target space theory is finite. Thus calculations based on this action will not suffer the nonrenormalisation problems of world volume actions in $D=11$ for example. 

Given the WV actions (5.15)and (5.17) the  next question to ask is what can we calculate? The natural quantities to calculate are scattering amplitudes and vacuum energy of the Target space theory. Proceeding by analogy with the 2d case reviewed in section III one could try to express the scattering amplitudes as topology changing processes with path integral representations:

\[
K(\Sigma _1  \oplus \Sigma _2 ;\Sigma _3  \oplus \Sigma _4 ) = \int\limits_{\Sigma _1  \oplus \Sigma _2 }^{\Sigma _3  \oplus \Sigma _4 } {DeD\omega D\chi e^{iS_{grav} } } 
\]
\hfill (5.20)

\noindent and the vacuum energy as the WV partition function:

\begin{eqnarray*}
 Z(WV)& =& \int {DNDN^i D\phi _1 } D\phi _2 D\phi _3 D\chi e^{iS_L }  \\ 
        & =& \int {DNDN^i DVDm_1 Dm_2 D\chi _1 D\chi _2 D\xi e^{iS_L } }  \\ 
 & =& \int {DA^{ + a} D\chi ^ +  } DA^{ - a} e^{i(S_{CS} (A^ +  ,\chi ^ +  ) - S_{CS} (A^ -  ))}  \\ 
             & =& \int {DeD\omega D\chi e^{iS_{grav} } }  \\ 
 \end{eqnarray*}
\hfill (5.21)

The initial and final two manifolds $\Sigma _1  \oplus \Sigma _2 $
and $\Sigma _3  \oplus \Sigma _4 $
can be thought of as a generalization of the  two pair of pants topology joined at the waist in string theory, only instead of circles at the boundary one has two dimensional manifolds. It is not clear that these manifolds have a limit of vanishing size where they have the effect of puncture or vertex operators as with the supermembrane in section IV or as with noncritical string theory [79].

Using these path integrals very explicit calculation of the topology changing amplitudes and partition functions can be found after imposing gauge fixing [80,81,82,83] and Faddev-Popov determinants. The most important result is that the path integrals using these 3d gravity actions are well defined and finite, related to the Reidemeister and Ray-Singer torsion and can be expressed as the integral over the modulii space of flat connections. There is a well known subtlety [82] associated with the range of integration for the lapse field $N$ in the above path integral. The issue is particularly important  here as the process of double dimensional reduction would link this path integral with the expressions from 2d gravity discussed in section III. Although the relation between the supermembrane approach and world volume 3d gravity approach is not clear, they both lead to a potentially finite description of M-theory in three Target space dimensions. Again this is because of the lack of transverse modes in the supermembrane case and the lack world volume gravitons in the WV 3d gravity formulation.

\section*{VI Conclusions}

The main interest in 3D M-theory is its potential to give a unified description of all the noncritical two dimensional string theories and to yield a finite target space theory of three dimensional gravity interacting with a single matter field. This differs from string theory in three dimensions which yields a target space theory of three dimensional gravity interacting with an infinite tower of matter fields which come from the transverse string oscillations. 

In this paper we have discussed several aspects of noncritical M-theory in three target dimensions. We have discussed the effective actions associated to noncritical M-theory as well as solutions to these effective actions including $AdS^3$, $AdS^2\times S^1$, and three dimensional black holes. We discussed the M2 supermembrane on $AdS^3$ including its action, currents, amplitudes and field equation for $T$ by relating it to a 3d sigma model over the supergroup $OSp(2|1;R)$. Finally we considered alternative world volume actions including the NSR spinning membrane and world volume pure 3d supergravity as a $OSp(2|1;R)\times SL(2,R)$ Chern-Simons theory that can be used to describe M-theory amplitudes as topology changing transitions on the world volume. 

The main results we found are the WV actions (4.1) (5.4), determination of the background metric (4.20) (5.9), the equation for the field $T$ (4.23) (5.11) , and the prescription for scattering amplitudes (4.28) (5.21) in the both the supermembrane and WV gravity approaches to 3D M-theory. 

It is useful to think of the two approaches as M-theoretic generalizations of the Green Schwarz [84] and Polyakov [47] treatment of string perturbation theory respectively. The process of double dimensional reduction reinforces this notion, as the supermembrane action yields the Green-Schwarz superstring on $AdS^2$ and the 3d WV gravity action leads to the induced 2d superliouville WS gravity action. The topological nature of the WV 3d gravity formulation as well as the fact that in some sense the theory is pure Liouville makes it more closely related to the topological gravity treatment of the $c=0$ string model in [85]. 

In the future it would be interesting to extend the continuous approaches to 3D M-theory of this paper using Discrete Light Cone Quantization on $AdS^3$ [86] or the random lattice description of WV 3d gravity [87]. This might connect with the more conventional M-atrix or dual gauge theory descriptions that have been useful in the eleven dimensional theory.

Another way to extend the 3D M-theory analysis is to noncritical F-theory in 2+2 dimensions. This theory would play the role of the 10+2 dimensional F-theory relation to M-theory in eleven dimensions [88]. Already for M-theory on $AdS^3$ an embedded description in flat 2+2 dimensions is a useful way to represent the operator algebra and effective actions [89]. Recent work on charged black holes with topology $AdS^2 \times S^2$, topological string models and large N two dimensional gauge theory [90] provide another instance of the utility of 2+2 to noncritical string models. Perhaps as a dual gauge description, large N two dimensional gauge theory may turn out to play the same role for noncritical M-theory as the large N four dimensional gauge models play in critical M-theory.

In summary the lack of transverse oscillations in 3D M-theory makes noncritical M- theory tractable. It also makes it difficult to extrapolate to it's 11D critical counterpart. However the existence of nontrivial solutions from  the 3D M-theory effective actions including three dimensional black holes indicate that there is much to be gained from the further understanding of three dimensional M-theory.

\section*{Acknowledgments}

I wish to thank M. Rocek, W. Siegel, P. van Nieuwenhuizen, and L. Smolin for useful comments and discussion.


\begin{thebibliography}{99}
\bibitem{McGreevy:2003kb} J.~McGreevy and H.~Verlinde,``Strings from tachyons: The c = 1 matrix reloaded,'' JHEP {\bf 0312}, 054 (2003) [arXiv:hep-th/0304224]. 
\bibitem{Douglas:2003up}
M.~R.~Douglas, I.~R.~Klebanov, D.~Kutasov, J.~Maldacena, E.~Martinec and N.~Seiberg,
``A new hat for the c = 1 matrix model,''
arXiv:hep-th/0307195.
\bibitem{McGreevy:2003ep}
J.~McGreevy, J.~Teschner and H.~Verlinde,
``Classical and quantum D-branes in 2D string theory,''
JHEP {\bf 0401}, 039 (2004)
[arXiv:hep-th/0305194].
\bibitem{Klebanov:2003km}
I.~R.~Klebanov, J.~Maldacena and N.~Seiberg,
``D-brane decay in two-dimensional string theory,''
JHEP {\bf 0307}, 045 (2003)
[arXiv:hep-th/0305159].
\bibitem{Davis:2004xb}
J.~Davis, L.~A.~Pando Zayas and D.~Vaman,
``On black hole thermodynamics of 2-D type 0A,''
JHEP {\bf 0403}, 007 (2004)
[arXiv:hep-th/0402152].
\bibitem{Danielsson:2004xf}
U.~H.~Danielsson, J.~P.~Gregory, M.~E.~Olsson, P.~Rajan and M.~Vonk,
``Type 0A 2D black hole thermodynamics and the deformed matrix model,''
JHEP {\bf 0404}, 065 (2004)
[arXiv:hep-th/0402192].
\bibitem{Gukov:2003yp}
S.~Gukov, T.~Takayanagi and N.~Toumbas,
``Flux backgrounds in 2D string theory,''
JHEP {\bf 0403}, 017 (2004)
[arXiv:hep-th/0312208].
\bibitem{McGreevy:2003dn}
J.~McGreevy, S.~Murthy and H.~Verlinde,
``Two-dimensional superstrings and the supersymmetric matrix model,''
JHEP {\bf 0404}, 015 (2004)
[arXiv:hep-th/0308105].
\bibitem{McGuigan:1991qp}
M.~D.~McGuigan, C.~R.~Nappi and S.~A.~Yost,
``Charged black holes in two-dimensional string theory,''
Nucl.\ Phys.\ B {\bf 375}, 421 (1992)
[arXiv:hep-th/9111038].
\bibitem{Mandal:2003tj}
G.~Mandal and S.~R.~Wadia,
``Rolling tachyon solution of two-dimensional string theory,''
JHEP {\bf 0405}, 038 (2004)
[arXiv:hep-th/0312192].
\bibitem{Verlinde:2004gt}
H.~Verlinde,
``Superstrings on AdS(2) and superconformal matrix quantum mechanics,''
arXiv:hep-th/0403024.
\bibitem{Witten:1991yr}
E.~Witten,
``On string theory and black holes,''
Phys.\ Rev.\ D {\bf 44}, 314 (1991).
\bibitem{Nakayama:2004vk}
Y.~Nakayama,
``Liouville field theory: A decade after the revolution,''
arXiv:hep-th/0402009.
\bibitem{Witten:1995ex}
E.~Witten,
``String theory dynamics in various dimensions,''
Nucl.\ Phys.\ B {\bf 443}, 85 (1995)
[arXiv:hep-th/9503124].
\bibitem{Acharya:2001gy}
B.~Acharya and E.~Witten,
``Chiral fermions from manifolds of G(2) holonomy,''
arXiv:hep-th/0109152.
\bibitem{Horava:1995qa}
P.~Horava and E.~Witten,
``Heterotic and type I string dynamics from eleven dimensions,''
Nucl.\ Phys.\ B {\bf 460}, 506 (1996)
[arXiv:hep-th/9510209].
\bibitem{Lukas:1997fg}
A.~Lukas, B.~A.~Ovrut and D.~Waldram,
``On the four-dimensional effective action of strongly coupled heterotic
string theory,''
Nucl.\ Phys.\ B {\bf 532}, 43 (1998)
[arXiv:hep-th/9710208].
\bibitem{Becker:2003wb}
M.~Becker, D.~Constantin, S.~J.~J.~Gates, W.~D.~.~Linch, W.~Merrell and J.~Phillips,
``M-theory on Spin(7) manifolds, fluxes and 3D, N = 1 supergravity,''
Nucl.\ Phys.\ B {\bf 683}, 67 (2004)
[arXiv:hep-th/0312040].
\bibitem{Duff:1996aw}
M.~J.~Duff,
``M theory (the theory formerly known as strings),''
Int.\ J.\ Mod.\ Phys.\ A {\bf 11}, 5623 (1996)
[arXiv:hep-th/9608117].
\bibitem{Li:1998vw}
M.~Li,
``Introduction to M theory,''
arXiv:hep-th/9811019.
\bibitem{Hughes:1986fa}
J.~Hughes, J.~Liu and J.~Polchinski,
``Supermembranes,''
Phys.\ Lett.\ B {\bf 180}, 370 (1986).
\bibitem{Bergshoeff:1987cm}
E.~Bergshoeff, E.~Sezgin and P.~K.~Townsend,
``Supermembranes And Eleven-Dimensional Supergravity,''
Phys.\ Lett.\ B {\bf 189}, 75 (1987).
\bibitem{Achucarro:1987nc}
A.~Achucarro, J.~M.~Evans, P.~K.~Townsend and D.~L.~Wiltshire,
``Super P-Branes,''
Phys.\ Lett.\ B {\bf 198}, 441 (1987).
\bibitem{Duff:1996zn}
M.~J.~Duff,
``Supermembranes,''
TASI lectures (1996)
[arXiv:hep-th/9611203].
\cite{Maldacena:1997re} \bibitem{Maldacena:1997re} J.~M.~Maldacena, ``The large N limit of superconformal field theories and supergravity,'' Adv.\ Theor.\ Math.\ Phys.\ {\bf 2}, 231 (1998) [Int.\ J.\ Theor.\ Phys.\ {\bf 38}, 1113 (1999)] [arXiv:hep-th/9711200]. 
\bibitem{Taylor:1999qk}
W.~I.~Taylor,
``The M(atrix) model of M-theory,''
arXiv:hep-th/0002016.
\bibitem{Taylor:2001vb}
W.~Taylor,
``M(atrix) theory: Matrix quantum mechanics as a fundamental theory,''
Rev.\ Mod.\ Phys.\  {\bf 73}, 419 (2001)
[arXiv:hep-th/0101126].
\bibitem{Duff:1987bx}
M.~J.~Duff, P.~S.~Howe, T.~Inami and K.~S.~Stelle,
``Superstrings In D = 10 From Supermembranes In D = 11,''
Phys.\ Lett.\ B {\bf 191}, 70 (1987).
\bibitem{Achucarro:1989dd}
A.~Achucarro, P.~Kapusta and K.~S.~Stelle,
``Strings From Membranes: The Origin Of Conformal Invariance,''
Phys.\ Lett.\ B {\bf 232}, 302 (1989).
\bibitem{Gross:1990ub}
D.~J.~Gross and I.~R.~Klebanov,
``One-Dimensional String Theory On A Circle,''
Nucl.\ Phys.\ B {\bf 344}, 475 (1990).
\bibitem{Achucarro:1992mb}
A.~Achucarro,
``Lineal gravity from planar gravity,''
Phys.\ Rev.\ Lett.\  {\bf 70}, 1037 (1993)
[arXiv:hep-th/9207108].
\bibitem{Achucarro:1993fd}
A.~Achucarro and M.~E.~Ortiz,
``Relating black holes in two-dimensions and three-dimensions,''
Phys.\ Rev.\ D {\bf 48}, 3600 (1993)
[arXiv:hep-th/9304068].
\bibitem{Cangemi:1992ri}
D.~Cangemi,
``One formulation for both lineal gravities through a dimensional reduction,''
Phys.\ Lett.\ B {\bf 297}, 261 (1992)
[arXiv:gr-qc/9207004].
\bibitem{Strominger:1998yg}
A.~Strominger,
``AdS(2) quantum gravity and string theory,''
JHEP {\bf 9901}, 007 (1999)
[arXiv:hep-th/9809027].
\bibitem{Spradlin:1999bn}
M.~Spradlin and A.~Strominger,
``Vacuum states for AdS(2) black holes,''
JHEP {\bf 9911}, 021 (1999)
[arXiv:hep-th/9904143].
\bibitem{Strominger:2003tm}
A.~Strominger,
``A matrix model for AdS(2),''
JHEP {\bf 0403}, 066 (2004)
[arXiv:hep-th/0312194].
\bibitem{Giveon:1993hm}
A.~Giveon, E.~Rabinovici and A.~A.~Tseytlin,
``Heterotic string solutions and coset conformal field theories,''
Nucl.\ Phys.\ B {\bf 409}, 339 (1993)
[arXiv:hep-th/9304155].
\bibitem{Banados:1992wn}
M.~Banados, C.~Teitelboim and J.~Zanelli,
``The Black hole in three-dimensional space-time,''
Phys.\ Rev.\ Lett.\  {\bf 69}, 1849 (1992)
[arXiv:hep-th/9204099].
\bibitem{Coussaert:1993jp}
O.~Coussaert and M.~Henneaux,
``Supersymmetry of the (2+1) black holes,''
Phys.\ Rev.\ Lett.\  {\bf 72}, 183 (1994)
[arXiv:hep-th/9310194].
\bibitem{Steif:1995zm}
A.~R.~Steif,
``Supergeometry of three-dimensional black holes,''
Phys.\ Rev.\ D {\bf 53}, 5521 (1996)
[arXiv:hep-th/9504012].
\bibitem{Coussaert:1994tu}
O.~Coussaert and M.~Henneaux,
``Self-dual solutions of 2+1 Einstein gravity with a negative  cosmological
constant,''
arXiv:hep-th/9407181.
\bibitem{Ayon-Beato:2004if}
E.~Ayon-Beato, C.~Martinez and J.~Zanelli,
``Birkhoff's theorem for three-dimensional AdS gravity,''
arXiv:hep-th/0403227.
\bibitem{Gegenberg:2003yz}
J.~Gegenberg and G.~Kunstatter,
``Using 3D stringy gravity to understand the Thurston conjecture,''
Class.\ Quant.\ Grav.\  {\bf 21}, 1197 (2004)
[arXiv:hep-th/0306279].
\bibitem{Carlip:1998uc}
S.~Carlip,
``Quantum gravity in 2+1 dimensions,'', Cambridge University Press (1998).
\bibitem{Siegel:1995hq}
W.~Siegel,
``Subcritical superstrings,''
Phys.\ Rev.\ D {\bf 52}, 3563 (1995)
[arXiv:hep-th/9503173].
\bibitem{Brink:1976sc}
L.~Brink, P.~Di Vecchia and P.~S.~Howe,
``A Locally Supersymmetric And Reparametrization Invariant Action For The
Spinning String,''
Phys.\ Lett.\ B {\bf 65}, 471 (1976).
\bibitem{Polyakov:1987ez}
A.~M.~Polyakov,
``Gauge Fields And Strings,'', Harwood Academic Publishers (1987).
\bibitem{Moore:1989yh}
G.~W.~Moore and N.~Seiberg,
``Taming The Conformal Zoo,''
Phys.\ Lett.\ B {\bf 220}, 422 (1989).
\bibitem{Dasgupta:2002iy}
A.~Dasgupta, H.~Nicolai and J.~Plefka,
``An introduction to the quantum supermembrane,''
Grav.\ Cosmol.\  {\bf 8}, 1 (2002)
[Rev.\ Mex.\ Fis.\  {\bf 49S1}, 1 (2003)]
[arXiv:hep-th/0201182].
\bibitem{Dasgupta:2000df}
A.~Dasgupta, H.~Nicolai and J.~Plefka,
``Vertex operators for the supermembrane,''
JHEP {\bf 0005}, 007 (2000)
[arXiv:hep-th/0003280].
\bibitem{deWit:1988ct}
B.~de Wit, M.~Luscher and H.~Nicolai,
``The Supermembrane Is Unstable,''
Nucl.\ Phys.\ B {\bf 320}, 135 (1989).
\bibitem{Nicolai:1998ic}
H.~Nicolai and R.~Helling,
``Supermembranes and M(atrix) theory,''
arXiv:hep-th/9809103.
\bibitem{Ezawa:1996fh}
K.~Ezawa and A.~Ishikawa,
``Osp(1$|$2) Chern-Simons gauge theory as 2D N = 1 induced supergravity,''
Phys.\ Rev.\ D {\bf 56}, 2362 (1997)
[arXiv:hep-th/9612031].
\bibitem{Park:1998un}
J.~Park and S.~J.~Rey,
``Green-Schwarz superstring on AdS(3) x S(3),''
JHEP {\bf 9901}, 001 (1999)
[arXiv:hep-th/9812062].
\bibitem{deWit:1998yu}
B.~de Wit, K.~Peeters, J.~Plefka and A.~Sevrin,
``The M-theory two-brane in AdS(4) x S(7) and AdS(7) x S(4),''
Phys.\ Lett.\ B {\bf 443}, 153 (1998)
[arXiv:hep-th/9808052].
\bibitem{Claus:1998fh}
P.~Claus,
``Super M-brane actions in AdS(4) x S(7) and AdS(7) x S(4),''
Phys.\ Rev.\ D {\bf 59}, 066003 (1999)
[arXiv:hep-th/9809045].
\bibitem{Dijkgraaf:1991ba}
R.~Dijkgraaf, H.~Verlinde and E.~Verlinde,
``String propagation in a black hole geometry,''
Nucl.\ Phys.\ B {\bf 371}, 269 (1992).
\bibitem{Giveon:1998ns}
A.~Giveon, D.~Kutasov and N.~Seiberg,
``Comments on string theory on AdS(3),''
Adv.\ Theor.\ Math.\ Phys.\  {\bf 2}, 733 (1998)
[arXiv:hep-th/9806194].
D.~Kutasov and N.~Seiberg,
``More comments on string theory on AdS(3),''
JHEP {\bf 9904}, 008 (1999)
[arXiv:hep-th/9903219].
Y.~Satoh,
``On string theory in AdS(3) backgrounds,''
arXiv:hep-th/0005169.
N.~Ishibashi, K.~Okuyama and Y.~Satoh,
``Path integral approach to string theory on AdS(3),''
Nucl.\ Phys.\ B {\bf 588}, 149 (2000)
[arXiv:hep-th/0005152].
Y.~Hikida, K.~Hosomichi and Y.~Sugawara,
``String theory on AdS(3) as discrete light-cone Liouville theory,''
Nucl.\ Phys.\ B {\bf 589}, 134 (2000)
[arXiv:hep-th/0005065].
D.~M.~Hofman and C.~A.~Nunez,
``Free field realization of superstring theory on AdS(3),''
JHEP {\bf 0407}, 019 (2004)
[arXiv:hep-th/0404214].
Y.~Hikida,
``String theory on Lorentzian AdS(3) in minisuperspace,''
JHEP {\bf 0404}, 025 (2004)
[arXiv:hep-th/0403081].
\bibitem{Berkovits:2002uc}
N.~Berkovits,
``Covariant quantization of the supermembrane,''
JHEP {\bf 0209}, 051 (2002)
[arXiv:hep-th/0201151].
\bibitem{Howe:1977hp}
P.~S.~Howe and R.~W.~Tucker,
``A Locally Supersymmetric And Reparametrization Invariant Action For A
Spinning Membrane,''
J.\ Phys.\ A {\bf 10}, L155 (1977).
P.~S.~Howe and R.~W.~Tucker,
J.\ Math.\ Phys.\  {\bf 19}, 869 (1978).
P.~S.~Howe and R.~W.~Tucker,
``Local Supersymmetry In (2+1)-Dimensions. 2. An Action For A Spinning
Membrane,''
J.\ Math.\ Phys.\  {\bf 19}, 981 (1978).
\bibitem{Sorokin:1999jx}
D.~P.~Sorokin,
``Superbranes and superembeddings,''
Phys.\ Rept.\  {\bf 329}, 1 (2000)
[arXiv:hep-th/9906142].
\bibitem{Bergshoeff:1988ui}
E.~Bergshoeff, E.~Sezgin and P.~K.~Townsend,
``On 'Spinning' Membrane Models,''
Phys.\ Lett.\ B {\bf 209}, 451 (1988).
\bibitem{Lindstrom:1988az}
U.~Lindstrom and M.~Rocek,
``A Super Weyl Invariant Spinning Membrane,''
Phys.\ Lett.\ B {\bf 218}, 207 (1989).
\bibitem{Kogan:1989ey}
Y.~I.~Kogan,
``The Off-Shell Closed Strings As The Topological Open Membranes: Dynamical
Transmutation Of World Sheet Dimension,''
Phys.\ Lett.\ B {\bf 231}, 377 (1989).
\bibitem{Livine:2002vq}
E.~R.~Livine and L.~Smolin,
``BRST quantization of matrix Chern-Simons theory,''
arXiv:hep-th/0212043.
L.~Smolin,
``The exceptional Jordan algebra and the matrix string,''
arXiv:hep-th/0104050.
L.~Smolin,
``M theory as a matrix extension of Chern-Simons theory,''
Nucl.\ Phys.\ B {\bf 591}, 227 (2000)
[arXiv:hep-th/0002009].
\bibitem{Dewitt:1967yk}
B.~S.~Dewitt,
``Quantum Theory Of Gravity. 1. The Canonical Theory,''
Phys.\ Rev.\  {\bf 160}, 1113 (1967).
\bibitem{Pioline:2002qz}
B.~Pioline and A.~Waldron,
``Quantum cosmology and conformal invariance,''
Phys.\ Rev.\ Lett.\  {\bf 90}, 031302 (2003)
[arXiv:hep-th/0209044].
\bibitem{Damour:2002et}
T.~Damour, M.~Henneaux and H.~Nicolai,
``Cosmological billiards,''
Class.\ Quant.\ Grav.\  {\bf 20}, R145 (2003)
[arXiv:hep-th/0212256].
\bibitem{Israel:2004vv}
D.~Israel, C.~Kounnas, D.~Orlando and P.~M.~Petropoulos,
``Electric / magnetic deformations of S**3 and AdS(3), and geometric cosets,''
arXiv:hep-th/0405213.
\bibitem{Mandal:1991tz}
G.~Mandal, A.~M.~Sengupta and S.~R.~Wadia,
``Classical solutions of two-dimensional string theory,''
Mod.\ Phys.\ Lett.\ A {\bf 6}, 1685 (1991).
\bibitem{Martinec:1984fs}
E.~J.~Martinec,
``Soluble Systems In Quantum Gravity,''
Phys.\ Rev.\ D {\bf 30}, 1198 (1984).
\bibitem{Matschull:1994cz}
H.~J.~Matschull,
``About loop states in supergravity,''
Class.\ Quant.\ Grav.\  {\bf 11}, 2395 (1994)
[arXiv:gr-qc/9403034].
B.~de Wit, H.~J.~Matschull and H.~Nicolai,
``Physical states in d = 3, N=2 supergravity,''
Phys.\ Lett.\ B {\bf 318}, 115 (1993)
[arXiv:gr-qc/9309006].
\bibitem{deAlfaro:1976je}
V.~de Alfaro, S.~Fubini and G.~Furlan,
``Conformal Invariance In Quantum Mechanics,''
Nuovo Cim.\ A {\bf 34}, 569 (1976).
\bibitem{Moncrief:1997pv}
V.~Moncrief and J.~E.~Nelson,
``Constants of motion and the conformal anti-de Sitter algebra in
(2+1)-dimensional gravity,''
Int.\ J.\ Mod.\ Phys.\ D {\bf 6}, 545 (1997)
[arXiv:gr-qc/9707032].
\bibitem{Rocek:1985bk}
M.~Rocek and P.~van Nieuwenhuizen,
``N >= 2 Supersymmetric Chern-Simons Terms As D = 3 Extended Conformal
Supergravity,''
Class.\ Quant.\ Grav.\  {\bf 3}, 43 (1986).
\bibitem{Achucarro:1987vz}
A.~Achucarro and P.~K.~Townsend,
``A Chern-Simons Action For Three-Dimensional Anti-De Sitter Supergravity
Theories,''
Phys.\ Lett.\ B {\bf 180}, 89 (1986).
\bibitem{Witten:1988hc}
E.~Witten,
``(2+1)-Dimensional Gravity As An Exactly Soluble System,''
Nucl.\ Phys.\ B {\bf 311}, 46 (1988).
\bibitem{Astorino:2002bj}
M.~Astorino, S.~Cacciatori, D.~Klemm and D.~Zanon,
``AdS(2) supergravity and superconformal quantum mechanics,''
Annals Phys.\  {\bf 304}, 128 (2003)
[arXiv:hep-th/0212096].
\bibitem{Moore:1991ir}
G.~W.~Moore, N.~Seiberg and M.~Staudacher,
``From loops to states in 2-D quantum gravity,''
Nucl.\ Phys.\ B {\bf 362}, 665 (1991).
\bibitem{Witten:1989sx}
E.~Witten,
``Topology Changing Amplitudes In (2+1)-Dimensional Gravity,''
Nucl.\ Phys.\ B {\bf 323}, 113 (1989).
\bibitem{Carlip:1994tt}
S.~Carlip and R.~Cosgrove,
``Topology change in (2+1)-dimensional gravity,''
J.\ Math.\ Phys.\  {\bf 35}, 5477 (1994)
[arXiv:gr-qc/9406006].
\bibitem{Carlip:1995jn}
S.~Carlip,
``A Phase space path integral for (2+1)-dimensional gravity,''
Class.\ Quant.\ Grav.\  {\bf 12}, 2201 (1995)
[arXiv:gr-qc/9504033].
\bibitem{Dasgupta:2001ue}
A.~Dasgupta and R.~Loll,
``A proper-time cure for the conformal sickness in quantum gravity,''
Nucl.\ Phys.\ B {\bf 606}, 357 (2001)
[arXiv:hep-th/0103186].
\bibitem{Carlip:1986cy}
S.~Carlip,
``Loop Calculations For The Green-Schwarz Superstring,''
Phys.\ Lett.\ B {\bf 186}, 141 (1987).
S.~Carlip,
Nucl.\ Phys.\ B {\bf 284}, 365 (1987).
\bibitem{Verlinde:1990ku}
E.~Verlinde and H.~Verlinde,
``A Solution Of Two-Dimensional Topological Quantum Gravity,''
Nucl.\ Phys.\ B {\bf 348}, 457 (1991).
\bibitem{Cho:1999fz}
~H.~Cho, T.~Lee and G.~W.~Semenoff,
``Two dimensional anti-de Sitter space and discrete light cone  quantization,''
Phys.\ Lett.\ B {\bf 468}, 52 (1999)
[arXiv:hep-th/9906078].
J.~H.~Cho and T.~Lee,
``Discrete light cone quantization of anti-de Sitter black hole and black
string,''
Nuovo Cim.\ B {\bf 115}, 991 (2000).
\bibitem{Ambjorn:1990ge}
J.~Ambjorn, B.~Durhuus and T.~Jonsson,
``Three-Dimensional Simplicial Quantum Gravity And Generalized Matrix Models,''
Mod.\ Phys.\ Lett.\ A {\bf 6}, 1133 (1991).
\bibitem{Vafa:1996xn}
C.~Vafa,
``Evidence for F-Theory,''
Nucl.\ Phys.\ B {\bf 469}, 403 (1996)
[arXiv:hep-th/9602022].
M.~Aganagic, R.~Dijkgraaf, A.~Klemm, M.~Marino and C.~Vafa,
``Topological strings and integrable hierarchies,''
arXiv:hep-th/0312085.
S.~Hewson and M.~Perry,
``The twelve dimensional super (2+2)-brane,''
Nucl.\ Phys.\ B {\bf 492}, 249 (1997)
[arXiv:hep-th/9612008].
D.~Kamani,
``pp-wave strings from membrane and from string in the spacetime with two  time
directions,''
Phys.\ Lett.\ B {\bf 564}, 123 (2003)
[arXiv:hep-th/0304236].
\bibitem{McKeon:2003xb}
D.~G.~C.~McKeon and C.~Schubert,
``Supersymmetry on AdS(3) and AdS(4),''
Class.\ Quant.\ Grav.\  {\bf 21}, 3337 (2004)
[arXiv:hep-th/0301225].
\bibitem{Vafa:2004qa}
C.~Vafa,
``Two dimensional Yang-Mills, black holes and topological strings,''
arXiv:hep-th/0406058.
\end{thebibliography}
\end{document}